\documentclass{aa}  

\usepackage{graphicx}
\usepackage{txfonts}
\usepackage{lipsum}
\usepackage{subcaption}         % necessary for continued figures, example in section 3
\usepackage{lscape}             % to rotate a single page table, example in appendix.
\usepackage{placeins}           % useful with \FloatBarrier, to keep 
\newcommand{\bra}[1]{\langle #1\rangle}

\newcommand{\nab}{\boldsymbol{\nabla}}
\newcommand{\AAA}{\boldsymbol{A}}
\newcommand{\BB}{\boldsymbol{B}}
\newcommand{\JJ}{\boldsymbol{J}}
\newcommand{\UU}{\boldsymbol{U}}
\newcommand{\SSSS}{\mbox{\boldmath ${\sf S}$} {}}
\newcommand{\Rey}{\mathrm{Re}}
\newcommand{\Rm}{\mathrm{Rm}}
\newcommand{\Pm}{\mathrm{Pm}}

\usepackage[export]{adjustbox}

\makeatletter
\AtBeginDocument{%
  \@ifpackageloaded{lineno}{%
    \nolinenumbers
    \let\linenumbers\relax
    
    \let\modulolinenumbers\relax
  }{}%
}
\makeatother

\usepackage{hyperref}
\begin{document}

\title{Imprints of primordial magnetic fields in gravitational collapse during early structure formation}

   \author{J. Schober\inst{1}\fnmsep\thanks{Corresponding author: schober@uni-bonn.de}
        \and M. Abramson\inst{2}
        \and S. Mandal\inst{2,3}
        \and S. Mtchedlidze\inst{4, 3, 5}
        \and T. Kahniashvili\inst{2, 3, 6}
        }
        
   \institute{Argelander-Institut f\"ur Astronomie, Universit\"at Bonn, Auf dem H\"ugel 71, 53121 Bonn, Germany
   \and McWilliams Center for Cosmology and Department of Physics, Carnegie Mellon University, Pittsburgh, PA 15213, USA
   \and School of Natural Sciences and Medicine, Ilia State University, 0194 Tbilisi, Georgia
   \and Dipartimento di Fisica e Astronomia, Universit\'a di Bologna, Via Gobetti 92/3, 40121, Bologna, Italy
   \and INAF Istituto di Radioastronomia, Via Gobetti 101, 40129 Bologna, Italy
   \and Department of Theoretical Astrophysics and Cosmology, Georgian National Astrophysical Observatory, GE-0179, Georgia}

   \date{Received September 30, 20XX}

  \abstract 
   {Primordial magnetic fields (PMFs) generated in the early Universe might have left observable imprints on present-day large-scale structure. However, the spatial scales on which primordial signatures are able to survive the nonlinear processes that accompany structure formation remain unclear. 
   }
   {The aim of this study is to investigate the evolution of the statistical properties of PMFs during the onset of gravitational collapse.} 
   {We performed a suite of high-resolution direct numerical simulations of 
   isothermal self-gravitating, magnetized gas clouds.
   By varying the viscosity, we probed different Reynolds-number regimes and follow the coupled evolution of gravitational collapse and magnetohydrodynamic turbulence.}
   {At sufficiently high Reynolds numbers, turbulence generated during collapse triggers the onset of a small-scale dynamo, which amplifies magnetic energy below the Jeans scale and modifies the magnetic energy spectrum significantly. 
   The question of whether dynamo amplification dominates the magnetic field evolution is determined by the competition between the dynamo growth time and the free-fall time.}
   {Our results highlight the importance of resolving the Jeans scale and the associated turbulent inertial range in cosmological magnetohydrodynamic (MHD) simulations to accurately capture the interplay between gravitational compression and dynamo amplification and to assess which structures retain memory of primordial fields.
}

   \keywords{Magnetic fields --
             Cosmology: theory --
             (Cosmology:) early Universe --
             (Cosmology:) large-scale structure of Universe --
            Magnetohydrodynamics (MHD) --
            Turbulence
}

\maketitle
\nolinenumbers

\section{Introduction}

Some of the most fundamental questions of modern physics could be answered by observing relics 
of the hot, dense plasma that permeated the early Universe.
One such relic that could have persisted to the present day is the primordial magnetic field (PMF) 
\citep{DN13,S16}. 
By placing observational constraints on PMFs, it may be possible to narrow the landscape of viable models for cosmic inflation \citep{TW88,Ratra1999}, 
baryogenesis \citep{Cornwall1997, GrassoRubinstein2001,KamadaLong2016,FujitaKamada2016}, 
and cosmological phase transitions 
\citep{V91,BBM96,TKBK12}.

To connect observational signatures of PMFs to early-Universe physics, it is essential to understand how PMFs evolve from their generation in the early Universe to the epoch of structure formation.
Their pre-recombination evolution is typically described within magnetohydrodynamics (MHD) \citep{BrandenburgEnqvistOlesen1996},
or, at very early times, chiral MHD 
\citep{Frohlich:2000e, BFR15, BSRKBFRK17, KamadaEtAl2023, SchoberEtAl2024}. 
In the absence of continued energy injection, PMFs decay while driving turbulence via the Lorentz force. 
In the helical case, magnetic helicity conservation leads to an inverse cascade, transferring magnetic energy from small to large scales \citep{Brandenburg2001}\footnote{In contrast, inverse cascading in nonhelical MHD turbulence was first observed in simulations \citep{ChristenssonEtAl2001} and has more recently been associated with the conservation of the Hosking integral \citep{HoskingSchekochihin21}.}. 
Around recombination, the decreasing ionization fraction weakens the coupling between magnetic fields and the gas, and turbulent decay becomes less efficient, although it does not cease entirely \citep{BanerjeeJedamzik2004}.

Overall, PMFs affect structure formation and the intergalactic medium through several well-established mechanisms, even after recombination.
Pre-recombination perturbations induced by PMFs are imprinted on the matter power spectrum
\citep{Wasserman:1978,Kim:1994zh,Subramanian:1997gi,Jedamzik:1996wp}.
This may affect the formation of dwarf galaxies \citep{SanatiEtAl2024}.
Below the magnetic Jeans scale, magnetic pressure suppresses baryonic fluctuations, and dissipation processes such as ambipolar diffusion heat the intergalactic medium \citep{Sethi:2004pe,Schleicher:2008aa}.
These effects leave imprints on high-redshift observables such as the Lyman-$\alpha$ forest and the redshifted $21\,$cm signal \citep{Pritchard:2011xb,Mesinger:2013nua}, which are commonly used to constrain the amplitude and spectral properties of PMFs \citep{Sethi:2004pe,SchleicherEtAl2009,Pandey:2014vga,Wagstaff:2015jaa}.
However, these high-redshift signatures of PMFs are partially degenerate with other effects on small-scale matter distribution, such as those induced by warm dark matter or massive neutrinos \citep{Viel:2013fqw}.

Low-redshift observables have also been intensively used to constrain the strength and structure of PMFs. 
Such observations include $\gamma$ rays from blazars and their secondary emission \citep{AharonianEtAl1994,LeeEtAl1995,DolagEtAl2004} (through which a lower bound of void magnetic fields of approximately $10^{-16}~\mathrm{G}$ has been inferred \cite{NeronovVovk2010}), 
synchrotron emission from galaxy clusters \cite{vanWeerenetal2019}, 
the so-called bridges of galaxy clusters \citep{BotteonEtAl2018,GovoniEtAl2019}, 
and from the stacked galaxy-group filaments \citep{VernstromEtAl2021}, 
and Faraday rotation measures (RM) within galaxy clusters \citep{,LoiEtal2026,KhadirEtAl2026,AlonsoLopezEtAl2026} and in the intergalactic medium \citep[see e.g., some of the recent work,][]{VernstromEtAl2019,OSullivanEtAl2020,CarrettiEtAl2023,MtchedlidzeEtAl2024,NeronovEtAl2024,CarrettiEtAl2025}.

Interpreting both low- and high-redshift observational constraints requires a consistent understanding of the intrinsic evolution of PMFs after recombination.
In particular, the evolution of the magnetic power spectrum during the first stages of gravitational collapse remains poorly understood.
\citet{MtchedlidzeEtAl2022} conducted simulations 
that incorporate various concepts of PMF generation, each characterized by different initial power spectra, as predicted by early-Universe generation mechanisms. 
These include scale-invariant spectra and those arising from phase-transition magnetogenesis, which are typically peaked at characteristic length scales determined by the underlying generation mechanism.
Their results demonstrate that the statistical properties of the large-scale RM signal can differ markedly between these scenarios, potentially offering a way to distinguish between them 
observationally \citep[see also][]{CarrettiEtAl2023,CarrettiEtAl2025,MtchedlidzeEtAl2024,MtchedlidzeEtAl2025arXiv,VazzaEtAl2025}.
Furthermore, these simulations reveal that gravitational collapse during structure formation drives a forward cascade of magnetic energy, effectively transferring magnetic power to smaller scales. 
This phenomenon manifests itself 
as a decrease in the characteristic correlation length of the magnetic field. 
An analytical model supporting this forward cascade behavior has also been proposed by \citet{AbramsonEtAl2025}, providing a physical interpretation for the simulation results.

At the same time, gravitational collapse naturally generates turbulence, which can amplify magnetic fields via the small-scale dynamo and modify the spectral transfer of magnetic energy in addition to the forward cascade.
The dynamo converts kinetic energy into magnetic energy through the stretching, twisting, and folding of magnetic field lines, leading to exponential amplification in the kinematic regime \citep{Kazantsev1968,KulsrudAnderson1992}.
Its growth rate increases with the hydrodynamic Reynolds number,
$\Rey = \sigma_v \ell_\mathrm{f}/(2\pi \nu)$,
where $\sigma_v$ is the velocity dispersion, $\ell_\mathrm{f}$ is the turbulent forcing scale, and $\nu$ is the viscosity \citep{SchoberEtAl2012,FederrathEtAl2011}.
This mechanism has been extensively studied in both idealized turbulence simulations and more complex environments, including isolated collapse, galaxies, and galaxy clusters \citep{SurEtAl2012,HigashiEtAl2024,RiederTeyssier2016,VazzaEtAl2018,DominguezEtAl2019,SteinwandelEtAl2021,MtchedlidzeEtAl_2023, PakmorEtAl2017,RiederTeyssier2017,Martin-AlvarezEtAl2018}.
Moreover, newly developed modeling techniques such as ``hyperdiffusion'' \citep{BiskampMueller2000,HaugenBrandenburg2004} and the use of ``supercomoving coordinates'' \citep{IrshadPEtAl2025, BrandenburgNtormousi2025} offer promising pathways toward improving our understanding of dynamo action and magnetic field amplification in realistic cosmological environments.

Despite this progress, the interplay between dynamo amplification and the spectral evolution of magnetic fields during collapse remains unclear. In particular, it is not yet understood how the small-scale dynamo interacts with the forward cascade of magnetic energy or how this interaction reshapes the magnetic power spectrum and potentially erases or modifies primordial signatures
on the scales of collapsing regions. Addressing this problem is challenging in cosmological simulations, where limited resolution and subgrid modeling restrict the ability to capture fully developed turbulence and high Reynolds numbers.

In this work, we investigate the evolution of PMFs during gravitational collapse using idealized direct numerical simulations (DNSs) of self-gravitating, magnetized gas.
By explicitly controlling the viscosity, and therefore the Reynolds number, we resolve the turbulent flow and study the onset of the small-scale dynamo in detail.
Our goals are twofold:
(i) to determine whether a forward cascade of magnetic energy emerges in a DNS framework and
(ii) to quantify how dynamo amplification influences the evolution of the magnetic power spectrum during the initial, approximately isothermal phase of collapse.
In contrast to cosmological simulations, which are limited in resolving fully developed turbulence, our approach isolates the magnetohydrodynamic processes during early collapse and is therefore most directly applicable to subhalo-scale gas dynamics during early structure formation.
Ultimately, this work aims to identify the characteristic length scales on which primordial magnetic field signatures may be modified or erased.

The paper is organized as follows.
In Sect.~\ref{sec_model} we introduce the governing equations and numerical methods.
Sect.~\ref{sec_DNS} presents the simulation results. 
In Sect.~\ref{sec_discussion} we compare our findings to previous numerical and analytical work. We summarize our conclusions in Sect.~\ref{sec_conclusion}.

\section{The model}
\label{sec_model}

Prior to the onset of nonlinear gravitational collapse, baryonic matter resides in mildly overdense regions that grow via gravitational instability within the cosmic web. 
As dark matter halos assemble, baryons fall into the associated potential wells and become compressed, eventually decoupling from the Hubble expansion. 
In many situations of interest, such as the formation of primordial star-forming minihalos or dense gas clouds within young galaxies, the central gas can be approximated as nearly isothermal over a substantial density range due to efficient radiative cooling \citep{Bromm2013}. 

In this regime, the collapse of a self-gravitating, pressure-confined sphere provides a useful idealized model of the nonlinear phase of structure formation on sub-halo scales. 
A supercritical Lane–Emden density profile
captures the transition from quasi-hydrostatic equilibrium to runaway gravitational collapse \citep{Shu1977}, while allowing controlled exploration of the turbulence generated during infall.

By construction, our setup neglects both the large-scale dark matter potential and cosmic expansion, and therefore does not aim to model the full cosmological environment.
The assumption of a static background is motivated by the fact that the collapse timescales considered here are short compared to the Hubble time, such that cosmic expansion has only a minor effect on the local dynamics.
This approach allows us to isolate the local gas dynamics, such as compressive amplification, turbulent energy injection, and magnetic cascade processes, which govern the evolution of PMFs once the gas has become self-gravitating and decoupled from the background flow.

\subsection{Governing equations}

To model the evolution of cosmic structures, we
consider an overdensity that collapses under self-gravity. 
The evolution of the mass density, $\rho$, is coupled with the
velocity field, $\UU$, and the gravitational potential, $\Phi$,
through the continuity equation,
\begin{eqnarray}
  \frac{\mathrm{D} \rho}{\mathrm{D} t} &=& - \rho \, \nab  \cdot \UU,
\label{eq_rho}
\end{eqnarray}
and the Poisson equation,
\begin{eqnarray}
  \nab^2 \Phi &=&  4 \pi G \rho,
\label{eq_Poisson}
\end{eqnarray}
where $G$ is the gravitational constant. Here, the evolution of the velocity field is governed by 
\begin{eqnarray}
  \rho \frac{\mathrm{D} \UU}{\mathrm{D} t} &=& \frac{1}{\mu_0} (\nab   \times   {\BB})  \times   \BB
     -\nab  P   - \rho\nab \Phi   + \nab  {\boldsymbol \cdot} (2\nu \rho \SSSS).
\label{eq_NS}
\end{eqnarray}
This equation describes how the momentum is determined by the Lorentz force, 
which includes the magnetic field, $\BB$, the gradients of the 
hydrodynamic pressure, $P$, and the gravitational potential, $\Phi$, while $\nu$ is the viscosity, and $\SSSS$ is the trace-free strain 
tensor with the components,
${\sf S}_{ij}=1/2(U_{i,j}+U_{j,i})-1/3~\delta_{ij} {\boldsymbol \nabla}
{\boldsymbol \cdot} \UU$.
The commas denote partial spatial derivatives 
and $\mathrm{D}/\mathrm{D} t = \partial/\partial t + \UU \cdot \nab$ is the
advective derivative.

When the term $-\nabla \Phi$ on the right-hand side of Eq.~(\ref{eq_NS}) dominates, a gravitational collapse sets in.
This gravitational collapse can, in principle, be affected by a magnetic field, 
for instance, by additional magnetic pressure. 
In this study, however, we restrict our attention to cases where magnetic fields are dynamically negligible and do not significantly modify the collapse. 
Nevertheless, we are interested in the evolution of the magnetic field,
$\BB$, during the collapse. 
To this end, we can solve the induction equation,
\begin{eqnarray}
  \frac{\partial \AAA}{\partial t} &=& 
    {\UU}  \times   {\BB}  - \eta \mu_0 \JJ, 
\label{eq_ind}
\end{eqnarray}
where the magnetic field is expressed via the 
vector potential as $\BB = \nabla \times \AAA$
and $\JJ=(\nabla \times \BB)/\mu_0$ is the electric current. 
Statistical properties of the magnetic field are encapsulated in the magnetic energy spectrum $E_\mathrm{M}$, which we normalize to get 
\begin{equation}
   \int E_\mathrm{M}(k,t)\,\mathrm{d} k \equiv \frac{\bra{\BB^2(t)}}{2\mu_0\rho_0} \equiv  \frac{B_\mathrm{rms}^2(t)}{2\mu_0\rho_0},
\label{eq_EM}
\end{equation}
where $\rho_0$ is the initial mean density.
In the following, we suppress the explicit time dependence for simplicity.
Different magnetogenesis scenarios produce different shapes of
$E_\mathrm{M}$, reflecting causality, energy injection, and turbulence. 
On the largest scales, the spectrum follows $E_\mathrm{M}(k) \propto k^4$,
consistent with causality constraints for primordial fields generated within the horizon. 
In the intermediate range, the spectrum transitions to a shallower slope, specifically $E_\mathrm{M}(k) \propto k^{-1}$,
representing a broad injection or coherence range. 
At small scales, the spectrum follows a Kolmogorov-type turbulent scaling $E_\mathrm{M}(k) \propto k^{-5/3}$,
corresponding to an inertial cascade toward dissipative scales, below which $E_\mathrm{M}(k)$ is exponentially suppressed.

\subsection{Phenomenology}

\subsubsection{Characteristic time and length scales}

The evolution of the magnetic field (growth or decay) in a self-gravitating cloud is governed by the interplay between amplification processes (i.e., gravitational collapse and dynamo action) and magnetic dissipation.
These processes are characterized by a set of fundamental time and length scales.

The timescale for collapse is the free-fall time,
\begin{eqnarray}
   t_{\text{ff}} = \sqrt{\frac{3\pi}{32G \rho_\mathrm{max}}},
\end{eqnarray}
which depends on the maximum density $\rho_\mathrm{max}$ of the system.  
Magnetic field amplification by turbulence occurs on the dynamo timescale,
$t_\mathrm{SSD}\equiv\gamma^{-1}$, where $\gamma$ is the small-scale dynamo growth rate, discussed later in this work.  

The relevant length scales are determined by gravitational stability and dissipation. 
The Jeans length,
\begin{eqnarray}
   r_{\text{J}} =  \sqrt{\frac{c_\mathrm{s}^2}{G \rho_\mathrm{max}}},
\end{eqnarray}
where $c_\mathrm{s}$ is the sound speed, marks the critical scale above which collapse proceeds.   
If turbulence arises solely from gravitational collapse, it is injected on a scale comparable to the Jeans length, $r_{\text{J}}$.
The turbulent cascade then transfers energy from $r_{\text{J}}$ to progressively smaller scales until it is dissipated 
at the viscous scale
\begin{eqnarray}
   \ell_\nu  = r_{\text{J}}\,\mathrm{Re}^{-3/4}.
\end{eqnarray}
Here, $\mathrm{Re}$ is the hydrodynamic Reynolds number\footnote{The factor of $2\pi$ results from a definition of the Reynolds number in wavenumber space.}, whereby
$\mathrm{Re}= \sigma_v r_{\text{J}}/(2\pi \nu)$ and the exponent of ${-3/4}$ is valid for incompressible turbulence. The interplay of these scales determines how the magnetic energy spectrum evolves during collapse; namely, whether amplification is dominated by turbulent dynamo action at small scales or whether the field is simply compressed with the gas.

\subsubsection{Two regimes of magnetic evolution}
\label{sec_regimes}

Depending on the relative ordering of $t_\mathrm{SSD}$ and $t_{\text{ff}}$, two different regimes emerge: 1) if $t_{\text{SSD}} \ll t_{\text{ff}}$, the small-scale dynamo operates efficiently, amplifying magnetic energy on the viscous scale well before the gas completes its collapse. In this regime, the growth of the magnetic spectrum is driven primarily by turbulence; 2) if $t_{\text{SSD}}  \gg t_{\text{ff}}$, the dynamo action is too slow to compete with the gravitational contraction. The magnetic field evolves passively, scaling with the gas density, and the spectrum largely mirrors the density distribution.

\subsubsection{Small-scale dynamo characteristics}
\label{sec_SSD}

The small-scale dynamo is a mechanism that converts turbulent kinetic energy into magnetic energy. 
It therefore depends intimately on the statistics of the velocity field. 
Traditionally, Kolmogorov-type turbulence has been studied, where the kinetic energy spectrum scales as $E_\mathrm{K} \propto k^{-5/3}$ in the inertial range. This type of turbulence is ideal because the dynamo seems to 
occur most naturally if the field lines are stretched, twisted, 
and folded by turbulent eddies \citep{Kazantsev1968}. 
However, the models can be extended to compressive turbulence, the limiting case of which would be Burgers turbulence, where $E_\mathrm{K}\propto k^{-2}$ \citep{SchoberEtAl2012, MartinsAfonsoEtAl2019}. 
For the onset of small-scale dynamo instability, the magnetic Reynolds number,
$\Rm=\sigma_v r_{\text{J}}/(2\pi \eta)$, where $\eta$ is the magnetic resistivity,
must exceed a critical value, $\Rm_\mathrm{c}$. 
For Kolmogorov turbulence, $\Rm_\mathrm{c}\approx100$, whereas for Burgers turbulence, the $\Rm_\mathrm{c}$ value is much higher \citep{HaugenEtAl2004}.

Once the dynamo is activated, it amplifies the magnetic field exponentially with the growth rate $\gamma$.
For the limit of large magnetic Prandtl numbers $\Pm=\Rm/\Rey$ the growth rate is predicted to be 
\begin{eqnarray}
   \gamma = c_1 \frac{\sigma_v}{\ell_\mathrm{f}} \Rey^{c_2}, 
\label{eq_gammaPmlarge}
\end{eqnarray}
where the ratio $\sigma_v / \ell_\mathrm{f}$ represents the eddy turnover rate at the forcing scale $\ell_\mathrm{f}$ with $\sigma_v$ denoting the velocity dispersion and 
$\ell_\mathrm{f} \approx r_\mathrm{J}$.
The constants $c_1$ and $c_2$ are different for different types of turbulence \citep{SchoberEtAl2012, BovinoEtAl2013}.
For Kolmogorov turbulence, $c_1 \approx 1$ and $c_2 = 1/2$, while for Burgers turbulence $c_1 \approx 0.2$ and $c_2 = 1/3$. These values are valid in the limit of large $\Pm$.
We note that numerical simulations typically operate in the regime $\Pm \approx 1$, due to the limited scale separation that can be achieved. 
The numerical solution of the Kazantsev equation by \citet{BovinoEtAl2013} indicates that, in this regime, $c_1$ takes a value similar to the one in the limit $\Pm \ll 1$. 
In \citet{SchoberEtAl2012b} the prefactors for low $\Pm$ have been found to be $c_1 \approx 0.03$ for Kolmogorov turbulence and $c_1 \approx 0.005$ for Burgers turbulence.
This will be important for the qualitative interpretation of the DNSs presented below.

While the eddy turnover time on the viscous wavenumber $k_\nu$ sets the growth rate of the magnetic field, 
the fastest buildup of magnetic energy occurs on the resistive wavenumber, 
\begin{eqnarray}
   k_\eta = k_\nu \Pm^{1/2}.
\label{eq_keta}
\end{eqnarray}
Once the magnetic field is strong enough to cause a backreaction on the velocity field,
the nonlinear phase begins, and the magnetic energy is shifted to larger spatial scales, potentially up to the forcing scale \citep{SchekochihinEtAl2002,SchleicherEtAl2013}, before saturation occurs \citep{SchoberEtAl2015}.

\vspace{-0.2cm}
\begin{table*}[t!]
\small
\caption{Overview of the key parameters of all runs presented in the main text of the paper.}
\centering
\begin{tabular}{ c | cccccccc | cc }
   \hline
Run   &  Res.      &  Helicity  &  $\begin{array}{c} E_\mathrm{M}(k) \\ \left[\mathrm{for}~k<k_*\right] \end{array}$  &  $k_*/k_\mathrm{J,0}$    &  $\begin{array}{c} E_\mathrm{M}(k) \\ \left[\mathrm{for}~k>k_*\right] \end{array}$  &  $\eta=\nu$  &  $B_\mathrm{rms}(0)$  &  $U_\mathrm{rms}(0)$  &  $\mathrm{max}(B_\mathrm{rms})$  &  $\mathrm{max}(U_\mathrm{rms})$  \\
\hline
H1a     &  $1152^3$  &  $1.0$       &  $\propto k^{-1}$    &  $3.0$      &  $\propto k^{-5/3}$  &  4.00e-02      &  $0.004102$          &  $0.899$             &  $0.005467$          &  $2.0716$            \\
H1b     &  $1152^3$  &  $1.0$       &  $\propto k^{-1}$    &  $3.0$      &  $\propto k^{-5/3}$  &  1.00e-03      &  $0.004102$          &  $0.899$             &  $0.02684$           &  $2.0685$            \\
H1c     &  $1152^3$  &  $1.0$       &  $\propto k^{-1}$    &  $3.0$      &  $\propto k^{-5/3}$  &  4.00e-04      &  $0.004102$          &  $0.899$             &  $0.0548$            &  $2.0521$            \\
N1a     &  $1152^3$  &  $0.0$       &  $\propto k^{-1}$    &  $3.0$      &  $\propto k^{-5/3}$  &  4.00e-02      &  $0.002831$          &  $0.899$             &  $0.0038$            &  $2.0671$            \\
N1b     &  $1152^3$  &  $0.0$       &  $\propto k^{-1}$    &  $3.0$      &  $\propto k^{-5/3}$  &  1.00e-03      &  $0.002831$          &  $0.899$             &  $0.01905$           &  $2.0864$            \\
H2b     &  $2304^3$  &  $1.0$       &  $\propto k^{-1}$    &  $3.0$      &  $\propto k^{-5/3}$  &  1.00e-03      &  $0.003934$          &  $0.8993$            &  $0.02277$           &  $1.879$             \\
H2b'  &  $2304^3$  &  $1.0$       &  $\propto k^{4}$     &  $100.0$  &  $\propto k^{-3}$    &  1.00e-03      &  $0.2023$            &  $0.8993$            &  $0.2023$            &  $1.9062$            \\
H2c     &  $2304^3$  &  $1.0$       &  $\propto k^{-1}$    &  $3.0$      &  $\propto k^{-5/3}$  &  4.00e-04      &  $0.003934$          &  $0.8993$            &  $0.03639$           &  $1.8746$            \\
H2d     &  $2304^3$  &  $1.0$       &  $\propto k^{-1}$    &  $3.0$      &  $\propto k^{-5/3}$  &  1.00e-04      &  $0.003934$          &  $0.8993$            &  $0.08068$           &  $1.7763$            \\
   \hline
\end{tabular}
\label{tab_DNSoverview}
\end{table*}

\subsection{Numerical simulations}
\label{subsec_DNSsetup}

We used the \textsc{Pencil Code} \citep{PencilCodeCollaboration2021} to simulate the evolution of an initial magnetic field with a given magnetic
energy spectrum during the collapse of an overdense region in space.
Specifically, we solved Eqs.~(\ref{eq_rho})-(\ref{eq_NS}) and (\ref{eq_ind}) on a three-dimensional (3D) grid. 
We employed the shock-dissipation scheme implemented in the \textsc{Pencil Code} to provide localized artificial diffusion in regions with strong gradients, thereby stabilizing the numerics and smoothing incipient shock structures. Here, the numerical domain has a size of $L^3$ and periodic boundary conditions.
The runs presented in the main part of the paper are characterized by a resolution
of $1152^3$ and $2304^3$. The dependence on numerical resolution is discussed in Appendix~\ref{sec_resolution}.

The length scales and wavenumbers reported in the DNS analysis are normalized in terms of the initial Jeans radius, $r_\mathrm{J,0}$, and the initial Jeans wavenumber, $k_\mathrm{J,0}=2\pi/r_\mathrm{J,0}$, respectively.
Density is normalized to the initial mean density, $\rho_0$, with $\rho_0 = 1$.
Time is normalized in terms of the initial free-fall time, $t_\mathrm{ff,0}$
and the velocities in terms of the sound speed, $c_\mathrm{s}$.
Finally, the magnetic field is reported in terms of the Alfv\`en velocity,
$B/\sqrt{4\pi \rho}$, which is again normalized by the sound speed, $c_\mathrm{s}$.

Our analysis is therefore presented in dimensionless units and should be interpreted in terms of the underlying control parameters of the problem. 
While our simulations include explicit viscosity and resistivity, these coefficients are prescribed for numerical control and do not represent realistic microphysical transport coefficients. 
In particular, we do not include additional nonideal MHD effects such as ambipolar diffusion, which can weaken the coupling between gas and magnetic field in low-density environments \citep[e.g.,][]{WhitworthEtAl2025}.

\subsubsection{Initial conditions}
\label{sec_IC}

For the initial density field, we adopted a spherically symmetric,
self-gravitating configuration corresponding to the isothermal
Lane-Emden solution. 
Assuming an isothermal equation of state,
$P = c_\mathrm{s}^2 \rho$, hydrostatic balance between pressure and
self-gravity yields the dimensionless equation,
\begin{eqnarray}
   \frac{1}{\xi^2} \frac{\mathrm{d}}{\mathrm{d}\xi} \left(
    \xi^2 \frac{\mathrm{d}\psi}{\mathrm{d}\xi}\right)
   = 
    e^{-\psi}.
\label{eq_LE}
\end{eqnarray}
The dimensionless potential $\psi$ is related to the density via
\begin{eqnarray}
   \rho = \rho_\mathrm{c} e^{-\psi},
\end{eqnarray}
where $\rho_\mathrm{c}$ is the central density
and the dimensionless radius, $\xi$, is connected to the physical
radius, $r$, via
\begin{eqnarray}
   r = \frac{c_\mathrm{s}}{\sqrt{4\pi G \rho_\mathrm{c}}}\,\xi.
\end{eqnarray}
The simulations are initialized with this density profile, centered in the cubic numerical domain. 
The profile is truncated at a fixed radius (approximately $0.3$ of the half-box size) and embedded in a constant-density background. 
In all simulations presented here, the central density exceeds the critical value for equilibrium, so that the configuration is gravitationally unstable and collapses under self-gravity.

The magnetic field is initialized with small fluctuations prescribed by a broken power law. 
In this work, we focus on intermediate length scales centered around the Jeans scale, spanning approximately one to two orders of magnitude both below and above it. We therefore choose the following initial condition:
\begin{equation}
  E_\mathrm{M}(k) = 
  \begin{aligned}
    \begin{cases} k^{-1} & \text{for~} k \leq 3, \\
      k^{-5/3} & \text{for~} k > 3 .
  \end{cases}
\end{aligned}
\label{eq_EM_IC}
\end{equation}

To provide a seed for dynamo action, we introduced decaying turbulent velocity fluctuations, with the peak of the energy spectrum chosen to coincide with the characteristic scale of gravitational instability. 
This initialization was required because the numerical setup cannot follow the gravitational collapse over sufficiently long timescales. In reality, we would expect turbulence to be generated self-consistently and rapidly as a natural consequence of the collapse.

\subsubsection{Overview of parameters}

Table~\ref{tab_DNSoverview} summarizes the set of simulation runs, listing their resolution, helicity, and initial magnetic energy spectra below and above the characteristic wavenumber, $k_\ast$. 
For each run, we report the magnetic diffusivity (which is set equal to the viscosity (i.e.,~$\eta = \nu$) such that $\Pm = 1$), the initial magnetic field strength and velocity field rms values ($B_\mathrm{rms}(0)$ and $U_\mathrm{rms}(0)$), 
and their maximum values attained during the simulations ($\text{max}(B_\mathrm{rms})$ and $\text{max}(U_\mathrm{rms})$).

\subsubsection{Analysis techniques}

\paragraph{Spatial averages}
  
When analyzing simulation data, we considered different types of spatial averages of a quantity $X$.
The root mean square value is defined as $X_\mathrm{rms} = \langle X^2 \rangle^{1/2}$. 
The total volume average is given by $\langle X \rangle$. 
In addition, we compute averages within the Jeans radius $\langle X \rangle_\mathrm{J}$, where the averaging is restricted to a sphere of radius $r_\mathrm{J}$ around the center of the numerical domain.

\paragraph{Power spectra}

$E_\mathrm{M}$ is the magnetic energy spectrum and has already been defined in Eq.~(\ref{eq_EM}).
Based on $E_\mathrm{M}$, we can define the correlation length of the magnetic field as
\begin{equation}
  k_\mathrm{M}^{-1} \equiv  \frac{\int k^{-1} E_\mathrm{M}(k,t)\,\mathrm{d} k }{\int E_\mathrm{M}(k,t)\,\mathrm{d} k}, 
\end{equation}
where the integration is performed over the entire numerical domain, namely, from $k=1$ up to
the maximally resolved wavenumber $k_\mathrm{max}$.

$E_\mathrm{K}$ is the kinetic energy spectrum, normalized such that
\begin{equation}
  \int E_\mathrm{K}(k,t)\,\mathrm{d} k \equiv \frac{\bra{\rho\UU^2}}{2}.    
\end{equation}
We define the correlation length of the velocity field as
\begin{equation}
  k_\mathrm{K}^{-1} \equiv  \frac{\int k^{-1} E_\mathrm{K}(k,t)\,\mathrm{d} k }{\int E_\mathrm{K}(k,t)\,\mathrm{d} k}. 
\end{equation}

Finally, $E_{\mathrm{log}\rho}$ is the density spectrum and normalized such that:
\begin{equation}
  \int E_{\mathrm{log}\rho}(k,t)\,\mathrm{d} k \equiv \bra{(\mathrm{log}\rho)^2} 
\end{equation}
and we define the correlation length of the logarithmic density as
\begin{equation}
  k_{\rho}^{-1} \equiv  \frac{\int k^{-1} E_{\mathrm{log}\rho}(k,t)\,\mathrm{d} k }{\int E_{\mathrm{log}\rho}(k,t)\,\mathrm{d} k}.
\end{equation}

\section{DNS results}
\label{sec_DNS}

\begin{figure*}
\centering
    \includegraphics[width=\textwidth]{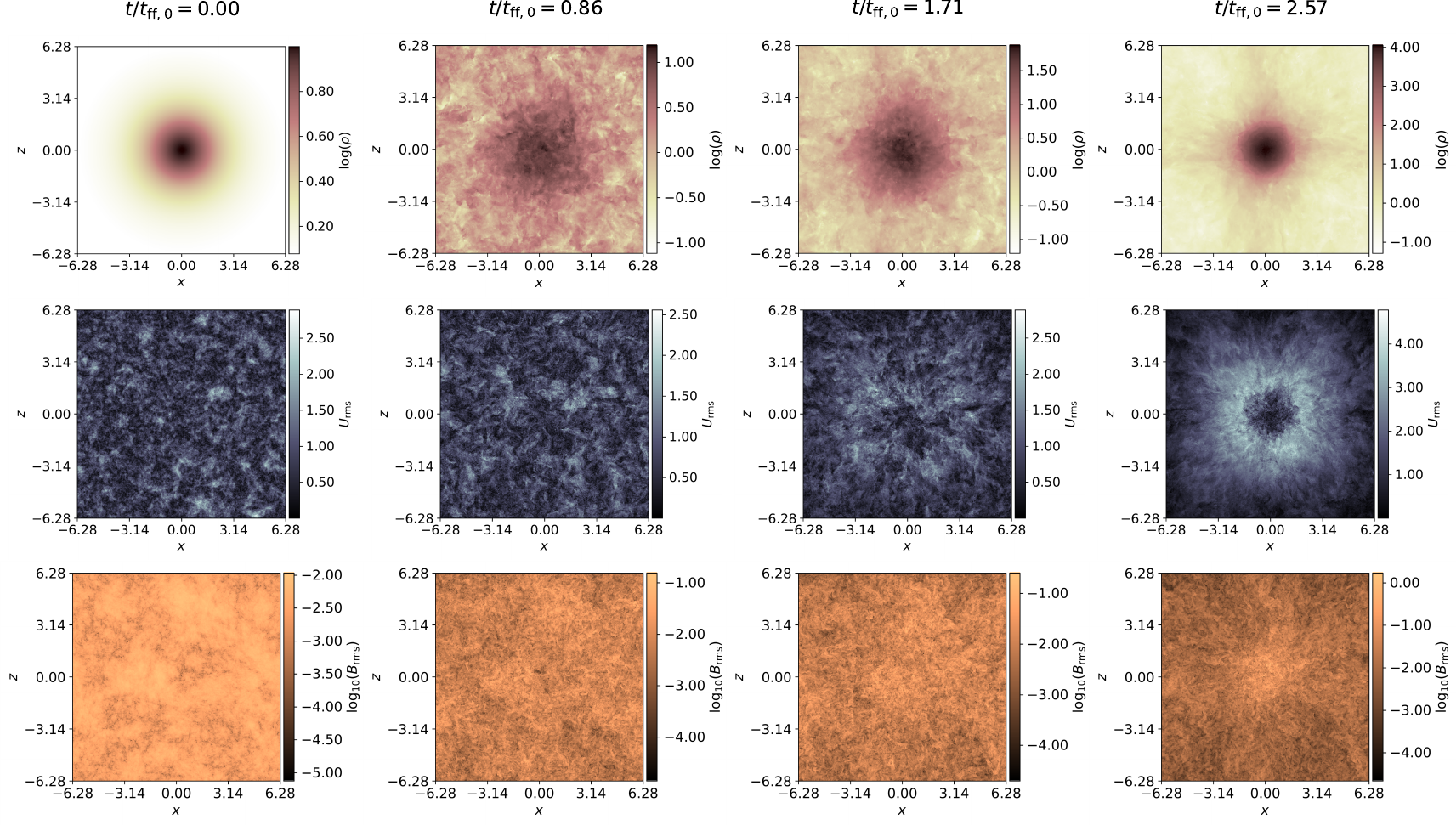}
\caption{Slices of the $x$--$z$ plane at $y=0$ of different quantities in Run H2b. Top: Logarithm of the density. Middle: rms velocity. Bottom: rms magnetic field.
Slices from left to right are taken at different
times: from the initial setup ($t/t_\mathrm{ff,0}=0$) to later times ($t/t_\mathrm{ff,0}=0.86, 1.71, 2.57$).}
\label{fig_slices_H2b}
\end{figure*}

\subsection{Reference run}

In this section, we take Run H2b as our reference case given it achieves a sufficiently
high $\Rey$ to potentially support small-scale dynamo action. 
At the same time, the inertial range remains resolved 
throughout the majority of the simulation. 
As with all runs, Run H2b crashes once the infall velocity 
becomes too large; namely, when it becomes larger than the speed of 
sound and shocks develop.
Therefore, we can only study the initial phase of the collapse with these DNSs.

\subsubsection{Collapse analysis}

Figure~\ref{fig_slices_H2b} presents slices of the simulation box from Run~H2b 
at various times, progressing from the initial conditions on the left to a snapshot near the final simulation time on the right.
Initially, the system is configured as an isothermal supercritical Lane-Emden density profile, as described in Sect.~\ref{sec_IC}.
By $t = 0.86~t_\mathrm{ff,0}$, where $t_\mathrm{ff,0}$ denotes the initial free-fall time of the system, the density distribution is significantly influenced by the turbulent velocity field initially imposed. 
At later times, however, gravitational collapse dominates, driving a rapid and continued increase in central density,
although the total density contrast remains limited compared to realistic protostellar collapse, where many orders of magnitude can be reached during the nearly isothermal phase \citep[e.g.,][]{OmukaiNishi1998}.

The evolution of the rms of the total velocity field is presented in the second row of Fig.~\ref{fig_slices_H2b}. 
The velocity field is initially set up with random fluctuations, but no artificial forcing is applied in the Navier-Stokes equation, as described in Sect.~\ref{subsec_DNSsetup}.
In the first half of the simulation, these random fluctuations are the dominant component of $U_\mathrm{rms}$. 
At $t \gtrsim 1.7~t_\mathrm{ff,0}$, the infall velocity dominates, as can be clearly seen in the slices. 
Note that the distance from the center, where $U_\mathrm{rms}$ reaches its spatial maximum, coincides with the radius where the spatial gradient of $\rho$ is highest (see upper panel of Fig.~\ref{fig_slices_H2b}), as expected from Eq.~(\ref{eq_NS}).

In the bottom row of Fig.~\ref{fig_slices_H2b}, the evolution of the rms of the magnetic field is shown. The initial fluctuations in Run~H2b are located mostly on the largest scale of the box. 
Later, $B_\mathrm{rms}$ is affected by both, the 
initial velocity fluctuations and 
the gravitational collapse, as the magnetic field is frozen in the gas. 
The magnetic field evolution will be discussed in more detail later in this paper.

\begin{figure}
\centering
    \includegraphics[width=0.4\textwidth]{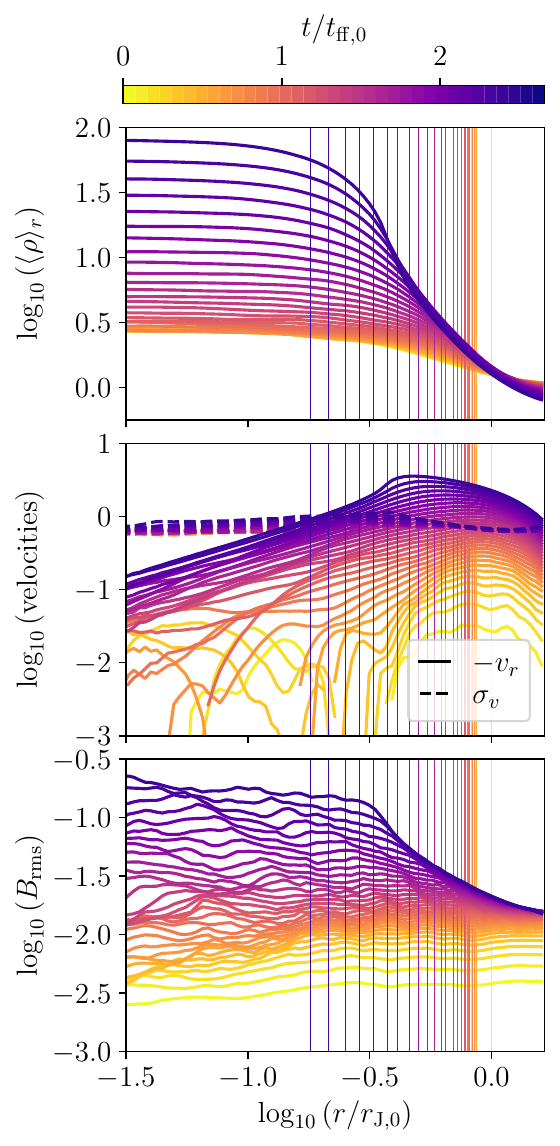}
\caption{Analysis of the evolution of radial profiles in Run H2b.
The panels show the radial profiles of the averaged density in a 
sphere with a radius $r$ (\textit{top}), radial velocity, $v_r$, and velocity dispersion, $\sigma_v$ (\textit{middle}), and the rms magnetic field strength (\textit{bottom}).
Thin vertical lines indicate the Jeans radius at different times.
Time is encoded by line color, as indicated by the color bar at the top of the figure.}
\label{fig_rhovB_r}
\end{figure}

A more quantitative analysis of the 
gravitational collapse in Run H2b is presented in 
Fig.~\ref{fig_rhovB_r}.
Here, the evolution of the radial profiles of various quantities is shown, where the radius $r$ is normalized by the initial Jeans radius $r_\mathrm{J,0}$.
The density, averaged within a radius, is given in the top panel, for different times as indicated by the color bar. 
The density increases fastest in the center, as expected.

In the middle panel of Fig.~\ref{fig_rhovB_r}, we show an analysis of the velocity field. 
We separate the radial component of the velocity field $v_r$ (averaged over the shell with radius $r$) and the velocity dispersion $\sigma_v$.
Due to the velocity fluctuations with which the simulation has been initialized throughout the 3D box, $\sigma_v$ already has a high value ($\mathrm{log}_{10} (\sigma_v) \approx -0.25$) at the beginning. 
The value of $\sigma_v$ is maintained more or less and is even slightly increasing up to ($\mathrm{log}_{10} (\sigma_v) \approx 0$) at 
the end of the simulation, due to the gravitational collapse. 
The radial velocity, on the other hand, is initially zero and increases strongly over time. 
The peak of $v_r(r)$ roughly follows the Jeans radius $r_\mathrm{J}$ which decreases during the collapse.
At the end of the simulation, the maximum value of $v_r$ is given by $\mathrm{log}_{10} (v_r) \approx 0.5$.

The magnetic field grows during the gravitational collapse, due to flux freezing in spherical compression, and potentially due to a small-scale dynamo. 
The radial dependence of $B_\mathrm{rms}$ is presented in the bottom panel of Fig.~\ref{fig_rhovB_r}. 
It can be seen that, until 
$t\approx 1.5~t_\mathrm{ff,0}$, $B_\mathrm{rms}$ grows at the same rate on all radii. 
Later, the growth of $B_\mathrm{rms}$ seems to occur primarily within the current Jeans radius, and $B_\mathrm{rms}$ at radii $r>r_\mathrm{J}(t)$ remains constant.

\begin{figure}
\centering
    \includegraphics[width=0.4\textwidth]{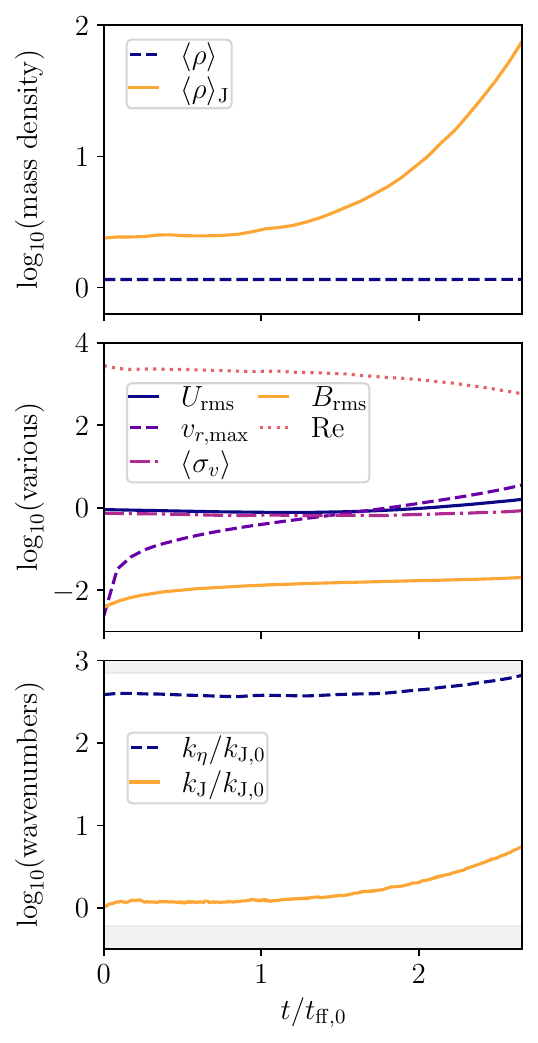}
\caption{Analysis of the time evolution of volume-averaged quantities in Run H2b.
Top: Mass density averaged over the 
entire simulation domain $\langle \rho \rangle$ (dashed blue line) and the density averaged 
over the instantaneous Jeans volume $\langle \rho \rangle_\mathrm{J}$ (solid orange line).
Middle: Quantities 
related to the velocity field, in particular, the rms velocity $U_\mathrm{rms}$ (solid blue line),
the maximum value of the radial velocity $v_{r,\mathrm{max}}$ (dashed purple line), and
the velocity dispersion averaged over the box $\langle\sigma_{v}\rangle$ (dashed-dotted magenta line). The rms magnetic field $B_\mathrm{rms}$ (solid orange line)
and the Reynolds number $\Rey$ (dotted red line) are also shown.
Bottom: Evolution of the resistive wavenumber $k_{\eta}$ (dashed blue line)
and the Jeans wavenumber, $k_{\mathrm{J}}$, are plotted.
Gray-shaded regions mark wavenumbers that lie outside the range captured by the resolution of the simulation.}
\label{fig_ts_collapse}
\end{figure}

Figure~\ref{fig_ts_collapse} shows the time evolution of several key quantities in Run~H2b.
The density averaged over the entire simulation domain, $\langle \rho \rangle$, remains constant throughout 
(as expected from mass conservation),
while the density averaged within the instantaneous Jeans volume, $\langle \rho \rangle_\mathrm{J}$, steadily increases — reflecting the localized collapse.
All velocity-related quantities grow over time, particularly during the later stages of the simulation.
The maximum radial velocity, $v_{r,\mathrm{max}}$, starts from zero and eventually surpasses 
both the root-mean-square velocity, $U_\mathrm{rms}$, and the velocity dispersion, $\langle \sigma_v \rangle$.
The root-mean-square magnetic field, $B_\mathrm{rms}$, also increases with time, although more slowly.
The Reynolds number has a high value of $\Rey \approx 3000$ initially, but decreases over the course of the simulation.
This decline is not due to a drop in $\langle \sigma_v \rangle$, as the turbulent velocity field is sustained by the collapse. 
The Reynolds number decreases rather due to an increase in the 
Jeans wavenumber, $k_{\mathrm{J}}$, which determines the scale 
at which turbulence is driven, as can be seen in the lower panel of 
Fig.~\ref{fig_ts_collapse}.
The same panel also shows that the dissipation scale, $k_\eta$, calculated from the evolving $\Rey$, remains more or less resolved throughout the entire duration of Run H2b 
(see the lower panel of Fig.~\ref{fig_ts_collapse}, where it is shown that $k_\eta/k_\mathrm{J,0}$ stays below the maximum resolved wavenumber until the end of the simulation).
As a result, the inertial range of turbulence (extending from $k_{\mathrm{J}}$ to $k_\eta$) is captured during the simulation.

\begin{figure}[t]
\centering
    \includegraphics[width=0.4\textwidth]{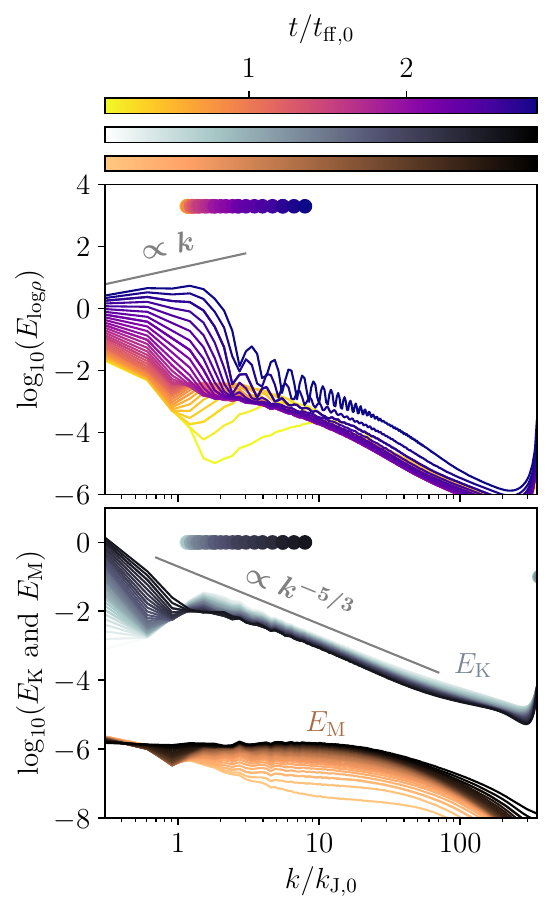}
\caption{Energy spectra for Run H2b.
Top: Density spectra. 
Bottom: Kinetic energy spectra (blue).
Magnetic energy spectra (red).
Solid circles indicate the (time-dependent) position of the Jeans wavenumber $k_\mathrm{J}$.
The initial viscous wavenumber, $k_\nu$, which coincides with the resistive wavenumber $k_\eta$, lies at the highest resolved wavenumber. 
Its location is marked by an open circle in the lower panel. Owing to the marginal resolution of $k_\nu$ and $k_\eta$ in this simulation, the marker is not fully visible.
}
\label{fig_spec_compact}
\end{figure}

The inertial range can be seen in the spectra presented in 
Fig.~\ref{fig_spec_compact}.
The kinetic energy spectrum $E_\mathrm{K}$ is plotted as blue lines in the lower panel.
Initially, $E_\mathrm{K}$ is peaked 
at the Jeans wavenumber and follows the 
Kolmogorov scaling, $\propto k^{-5/3}$.
This scaling remains the same throughout the simulations. 
However, $E_\mathrm{K}$ increases strongly at $k<k_{\mathrm{J},0}$ due to 
the gravitational collapse that generates radial inflows toward the center of the domain. 
The magnetic energy spectrum $E_\mathrm{M}$ is presented in the same panel as the orange lines and will be discussed in the next section. 
For completeness, the top panel of Fig.~\ref{fig_spec_compact} displays the 
density spectrum, $E_{\log \rho}$.
At low wavenumbers ($k < k_{\mathrm{J},0}$), we observe the emergence of a linear scaling with $k$, while at higher wavenumbers ($k > k_{\mathrm{J},0}$), the spectrum exhibits oscillatory features (``wiggles'') that have also been reported in \citet{BrandenburgNtormousi2022}.

\subsubsection{Magnetic field evolution}

During the collapse, the rms magnetic field strength increases by approximately a factor of $10$ in Run H2b, 
as shown in the middle panel of Fig.~\ref{fig_ts_collapse}.
The amplification initially proceeds approximately uniformly throughout the computational domain.
In the final stage, however, 
$B_\mathrm{rms}$ increases by an additional order of magnitude within the innermost final Jeans scale.
Concurrently, the magnetic field structure changes, as we see from the evolution of $E_\mathrm{M}$, which is shown
in the lower panel of Fig.~\ref{fig_spec_compact}.
In the remainder of this subsection, we investigate the mechanisms 
driving this amplification — specifically, whether it is primarily 
due to gravitational compression or the action of a small-scale dynamo.

In Sect.~\ref{sec_regimes}, we argue that a small-scale dynamo dominates
magnetic field amplification during gravitational collapse 
if its inverse growth rate is shorter than the free-fall time, $t_\mathrm{ff}$.
To check whether this condition is met in Run H2b, 
we compare the $t_\mathrm{ff}$, which evolves over time, with the inverse growth rate $\gamma^{-1}$ expected from dynamo theory, as discussed in Sect.~\ref{sec_SSD}.
The growth rate in the limit of large $\Pm$ is 
given in Eq.~(\ref{eq_gammaPmlarge}).
However, our simulations have been performed for $\Pm=1$ due to numerical limitations.
For $\Pm=1$, the growth rate of the small-scale dynamo 
instability~(\ref{eq_gammaPmlarge}) measured in DNSs agrees usually 
better when using prefactors, $c_1$, 
derived for the limit of small $\Pm$ \citep[see e.g.,][]{KrielEtAl2025}.
In \citet{SchoberEtAl2012b} these prefactors have been 
found to be $c_1\approx0.03$ for Kolmogorov turbulence and 
$c_1\approx0.005$ for Burgers turbulence, and we will use these values for the following 
comparison with our simulations.

\begin{figure}[t]
\centering
    \includegraphics[width=0.4\textwidth]{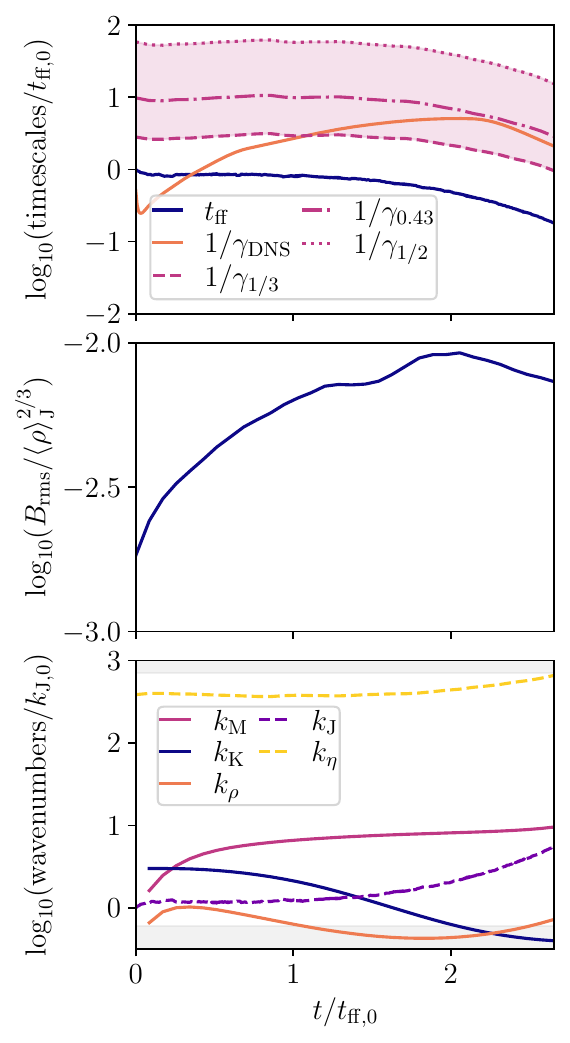}
\caption{Analysis of quantities related to the dynamo instability for Run H2b. 
Top: Evolution of the free-fall time is compared to the inverse of the measured growth rate of $B_\mathrm{rms}$ ($\gamma_\mathrm{DNS}$) and different theoretically predicted growth rates ($\gamma_{1/3}$, $\gamma_{0.43}$, and $\gamma_{1/2}$) in the red band. 
Middle: Evolution of $B_\mathrm{rms}/\langle \rho \rangle^{2/3}$.
Bottom: Various characteristic wavenumbers: the correlation wavenumbers of the magnetic ($k_\mathrm{M}$), the velocity ($k_\mathrm{K}$), the density field ($k_\rho$), as well as the Jeans wavenumber ($k_\mathrm{J}$) and the resistive wavenumber ($k_\eta$).}
\label{fig_growth_t}
\end{figure}

In the top panel of Fig.~\ref{fig_growth_t}, different theoretical predictions\footnote{Technically, in the limit of small Prandtl numbers, $\gamma$ depends on the magnetic Reynolds number $\Rm$. However, we consider $\Pm=1$ where $\Rey=\Rm$.}
for the dynamo growth rate are presented: Kolmogorov turbulence where $\gamma=\gamma_{1/3} = 0.03\, \Rey^{1/2}$, 
and turbulence produced by solenoidal forcing \citep{FederrathEtAl2010}
where $\gamma=\gamma_{0.43} = 0.019 \,\Rey^{0.4}$,
and Burgers turbulence where $\gamma=\gamma_{1/2} = 0.005\, \Rey^{1/3}$\footnote{The indices in $\gamma$ correspond to the scaling of the velocity fluctuations $u_\ell$ with scales $\ell$. 
In Kolmogorov turbulence, $u_\ell \propto \ell^{1/3}$, while in Burgers turbulence $u_\ell \propto \ell^{1/2}$; see e.g.,~ \citet{SchoberEtAl2012}.}.
All of these inverse growth rates, $\gamma_{1/3}^{-1}$, $\gamma_{0.43}^{-1}$ and $\gamma_{1/2}^{-1}$ are larger than the free-fall time $t_\mathrm{ff}$ during the entire time simulated in Run H2b.
With this and the fact that the Reynolds number is above the critical one (see Fig.~\ref{fig_ts_collapse}), a dynamo could be possible but slower than the amplification of the magnetic field via gravitational collapse.

The measured inverse growth rate of $B_\mathrm{rms}$, $\gamma_\mathrm{DNS}^{-1}$, is also presented in the top panel of Fig.~\ref{fig_growth_t}. 
As the simulation is set up with initial velocity fluctuations that initially decay, 
$\gamma_\mathrm{DNS}^{-1}$ increases until, at $t/t_\mathrm{ff,0}\approx 2$, the gravitational collapse produces more turbulence, and the time scale of the small-scale dynamo decreases again. 
Note that $t/t_\mathrm{ff,0}\approx 2$ coincides with the moment when $U_\mathrm{rms}$ starts to increase; see the middle panel of Fig.~\ref{fig_ts_collapse}.
With the kinetic energy spectrum $E_\mathrm{K}$ maintaining a
Kolmogorov-like scaling $\propto k^{-5/3}$ throughout the simulation (see Fig.~\ref{fig_spec_compact}), 
the comparison with $\gamma_{1/3}^{-1}$ seems appropriate. 
However, at late times, we expect more and more compressive modes, as turbulence is driven through the gravitational collapse. 
Overall, $\gamma_\mathrm{DNS}$ agrees best with $\gamma_{0.43}$ obtained from Kazantsev theory with the characteristics of a velocity field found in simulations of solenoidal forcing.

As $\gamma^{-1}$ is larger than $t_\mathrm{ff}$, Run H2b is not in the regime where the small-scale-dynamo dominates the magnetic field amplification. 
However, this run has the highest ratio $\gamma^{-1}/t_\mathrm{ff}$, for which the resistive scale remains resolved throughout the simulation. 
For this reason, we adopted it as the reference run.
The analysis shown in the top panel of Fig.~\ref{fig_growth_t} is also 
presented for selected other runs in Fig.~\ref{fig_timescales_t} in the appendix. 
In the appendix, Run H2d is also shown, which has the highest $\Rey$ in our suite of runs. 
For Run H2d  $\gamma^{-1}/t_\mathrm{ff}$ is smaller, and the dynamo should be more pronounced. However, the resistive scale in this run is not resolved and the dynamo cannot perform at its best. 

Still, even in H2b, the measured $\gamma_\mathrm{DNS}$ falls within the range of theoretical predictions as discussed in Sect.~\ref{sec_SSD}, indicating small-scale dynamo action. 
This interpretation is further supported by the time evolution of $B_\mathrm{rms}/\langle\rho\rangle_\mathrm{J}^{2/3}$.
This ratio would remain constant if the magnetic field is only increased due to
gravitational compression (or even decrease due to dissipation of magnetic energy).
However, in the middle panel of Fig.~\ref{fig_growth_t},
it is shown that $B_\mathrm{rms}/\langle\rho\rangle_\mathrm{J}^{2/3}$ 
increases by roughly a factor of two. 
This points towards dynamo action. 
The time evolution of some characteristic wavenumbers is presented in the bottom panel of Fig.~\ref{fig_growth_t}. 
Although the correlation wavenumber of $E_\mathrm{M}$ increases slightly, it never reaches its peak on $k_\eta$, which would be expected during the kinematic phase of the dynamo.
Again, we only observe the onset of the dynamo; the simulation crashes before some of the dynamo characteristics could unfold. 
It is also interesting to notice that most of the magnetic energy resides on scales smaller than coherence scales of velocity and density fields; similar results have also been found in cosmological simulations of \citet{MtchedlidzeEtAl_2023} and in MHD simulations of \citet{ChoRyu2009}.

\begin{figure}
\centering
    \includegraphics[width=0.4\textwidth]{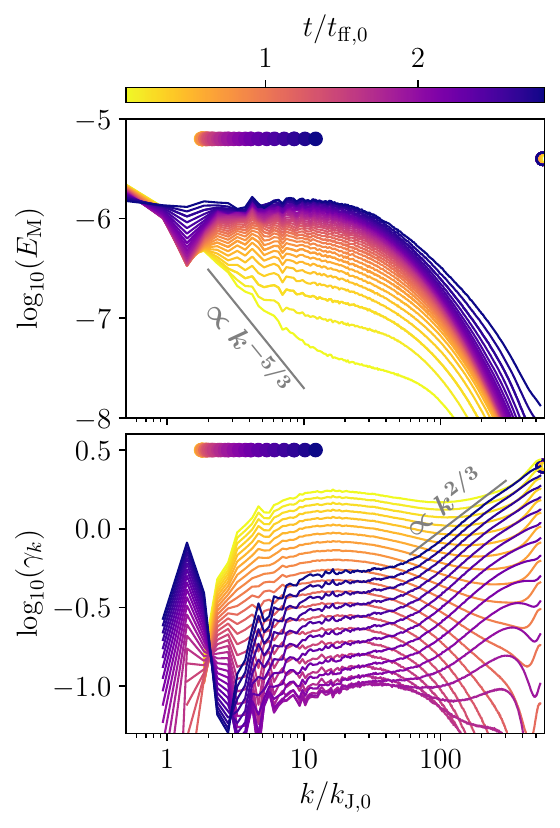}
\caption{Analysis of magnetic energy growth in Run H2b.
Top: Time evolution of the magnetic energy spectrum, $E_\mathrm{M}$, as indicated by the colorbar.
Bottom: Scale-dependent growth rate, $\gamma_k$.
Solid circles indicate the (time-dependent) position of the Jeans wavenumber $k_\mathrm{J}$ 
and open circles indicate the (time-dependent) position of the resistive wavenumber $k_\eta$.
}
\label{fig_spec_gamma}
\end{figure}

The top panel of Fig.~\ref{fig_spec_gamma} shows the evolution 
of the magnetic energy spectrum for Run H2b.
Initially, $E_\mathrm{M}$ follows a power-law scaling of 
$k^{-5/3}$ for wavenumbers $k > k_* = 3\, k_\mathrm{J,0}$, but this scaling quickly 
breaks down as magnetic energy grows across most scales during the 
collapse; the exception is at the largest length scale (i.e.,~the size of the numerical domain).
Here, $E_\mathrm{M}$ decreases slightly. 
This decrease in magnetic energy on large scales may be a numerical artifact. 
However, a similar decline in the peak of $E_\mathrm{M}$ during gravitational collapse in large-scale structure formation has been reported by \citet{MtchedlidzeEtAl2022}.
Overall, in Run H2b, the evolution of $E_\mathrm{M}$ exhibits a forward cascade, with the magnetic energy tracking the evolution of the Jeans wavenumber $k_\mathrm{J}$.
The position of $k_\mathrm{J}$ therefore appears to set the characteristic scale of the magnetic field during the collapse.
Interestingly, when the initial maximum of $E_\mathrm{M}$ is located at $k > k_\mathrm{J,0}$, as in Run H2b', we instead observe a transfer of magnetic energy toward smaller $k$, with the spectrum eventually peaking at $k_\mathrm{J}$. This inverse cascade in Run H2b' is discussed in Appendix~\ref{sec_H2bd}.

Notably, we do not observe significant growth in $E_\mathrm{M}$ in Run H2b at 
the dissipation scale ($k = k_\eta$, indicated by the open circles in the figure), where amplification would be expected if a small-scale dynamo were active. 
This could be due to the initially very low magnetic energy at these scales, delaying the onset of dynamo action.
By $t \gtrsim 2\,t_\mathrm{ff,0}$; however, some amplification begins to appear at high wavenumbers. 
The amplification at high wavenumbers is more clearly illustrated in the bottom panel of Fig.~\ref{fig_spec_gamma}, which shows the magnetic energy growth rate per mode, $\gamma_k(k)$, for Run~H2b.
If a small-scale dynamo is operating, it amplifies the magnetic energy 
at a given scale, $k$, on the timescale of the local eddy turnover time, such that
\begin{eqnarray}
   \gamma_k \approx u_k k \propto k^{2/3},
\label{eq_k23}
\end{eqnarray}
for Kolmogorov turbulence, where $u_k \propto k^{-1/3}$.
This characteristic scaling behavior becomes visible in the 
lower panel of Fig.~\ref{fig_spec_gamma} at 
$t \gtrsim 2t_\mathrm{ff,0}$, suggesting the onset of small-scale dynamo 
activity at later stages of collapse. 

Overall, the analysis of Run H2b suggests that, in addition to gravitational compression, the magnetic field 
is also amplified by a small-scale dynamo. 
More evidence to support the presence of an additional dynamo in Run H2b is presented in Appendix~\ref{sec_work_terms}, where we analyze the work terms in the flow,
and in Appendix~\ref{sec_correlations}, where we examine the spatial correlations between $B_\mathrm{rms}$ and vorticity.

\subsection{Transition to a dynamo regime}

\begin{figure}[h!]
\centering
    \includegraphics[width=0.4\textwidth]{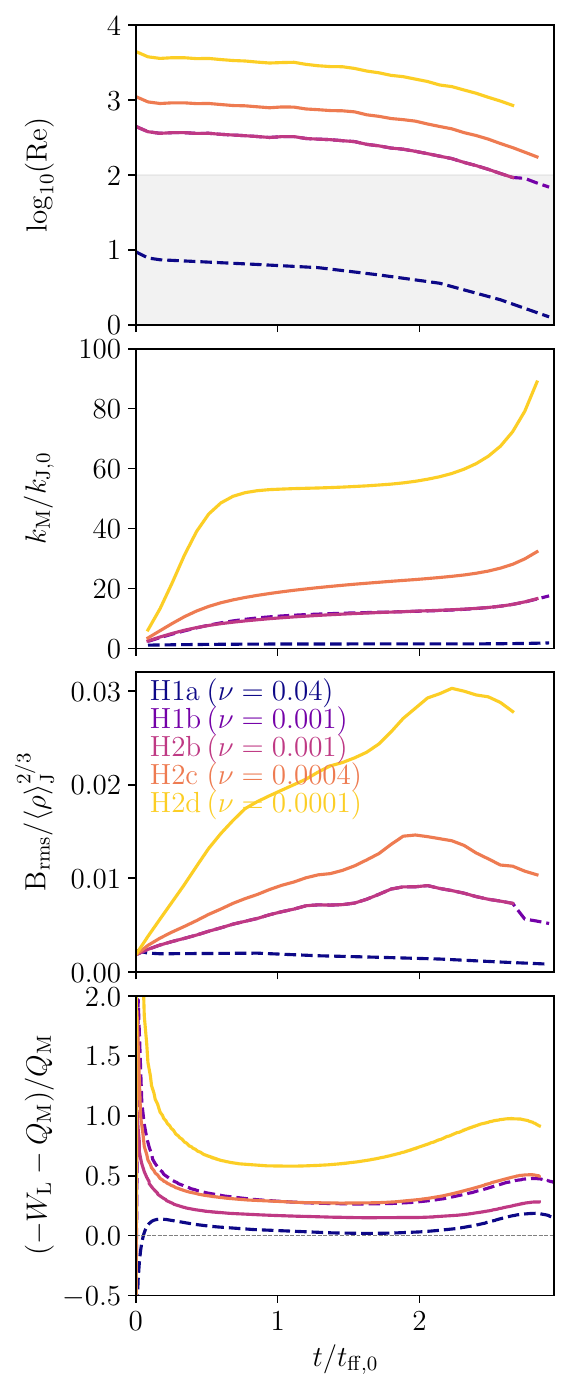}
\caption{Analysis of magnetic field growth in simulations with different viscosity $\nu$.
The different line colors refer to different runs, as indicated on the legend shown in the third panel. 
Different line styles indicate the resolution: $1152^3$ (dashed lines) and $2304^3$ (solid lines).
From top to bottom: Reynolds number (\textit{top}), ratio of the integral wave number of the magnetic energy spectrum $k_\mathrm{M}$ normalized by $k_{\mathrm{J},0}$ (\textit{second panel}), $B_\mathrm{rms}/\langle\rho\rangle_{\mathrm{J}}^{2/3}$ (\textit{third panel}), and  
$(-W_\mathrm{L} - Q_\mathrm{M})/Q_\mathrm{M}$ (\textit{bottom}). 
The shaded region in the top panel marks
$\Rey<100$, where the magnetic Reynolds number is below the critical threshold for small-scale dynamo action.
}
\label{fig_growth_eta}
\end{figure}

Next, we turn to the question of how magnetic field amplification in a self-gravitating collapsing cloud is influenced by the level of turbulence in the system. For this purpose, we conducted a parameter study by varying the viscosity $\nu$ (and likewise the magnetic resistivity, $\eta$) in our simulations.

\begin{figure*}
\centering
    \includegraphics[width=\textwidth]{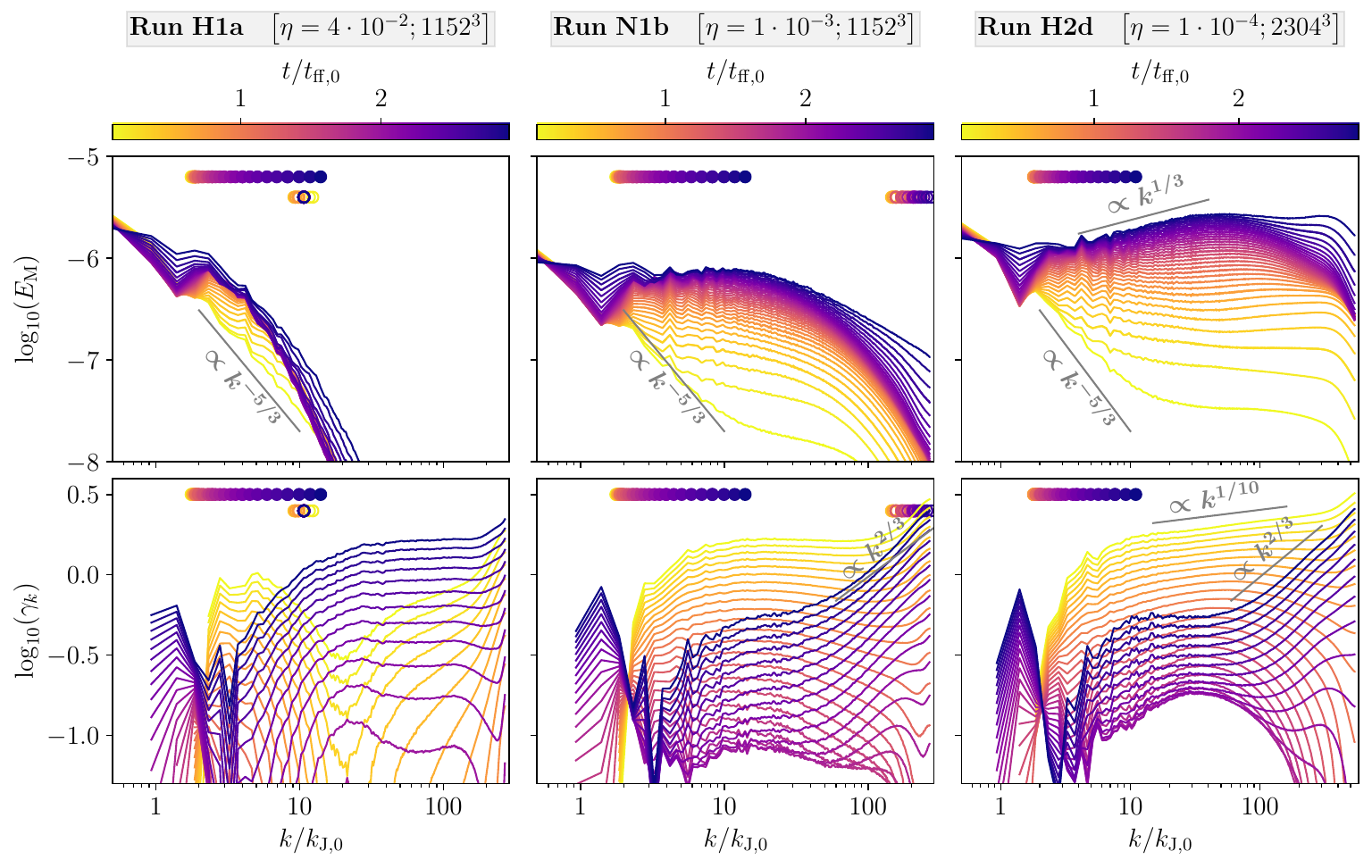}
\caption{Comparison of the spectral evolution of magnetic energy in different runs.
From left to right, the value of $\eta$ decreases. Specifically, we show Run H1a ($\eta = 4 \times 10^{-2}$), Run N1b ($\eta = 10^{-3}$), and Run H2d ($\eta = 10^{-4}$).
Top and bottom panels: Same analysis as done for H2b in Fig.~\ref{fig_spec_gamma}.
}
\label{fig_spec_eta}
\end{figure*}

In Fig.~\ref{fig_growth_eta}, we compare simulations with different $\nu$ values and, hence, with 
different Reynolds numbers, $\Rey$.
The top panel of Fig.~\ref{fig_growth_eta} shows the time evolution of $\Rey$ for different runs.
The highest $\Rey \approx 4000$ is reached in Run H2d, although 
the viscous and resistive scales, $k_\nu = k_\eta$, are not resolved in this case.
Our reference run, Run H2b, has the highest $\Rey \approx 400$, while $k_\eta$ remains resolved until the simulation is over.
Run H1a starts with the lowest $\Rey \approx 10$, well below the critical value of $\Rey \approx 100$ required for small-scale dynamo action \citep{HaugenEtAl2004}.
Generally, we find that $\Rey$ decreases over time in all runs throughout the simulated period. 
This occurs despite initializing the simulations with velocity fluctuations that are later sustained by turbulence driven during gravitational collapse, keeping $U_\mathrm{rms}$ approximately constant. 
However, the turbulent driving scale, set by the evolving Jeans scale, decreases over time, leading to the observed decrease 
in $\Rey$.

The second panel of Fig.~\ref{fig_growth_eta} shows the evolution of the magnetic energy integral scale, $k_\mathrm{M}$.
As expected for a small-scale dynamo, $k_\mathrm{M}$ increases faster with decreasing $\nu$, namely, with increasing $\Rey$.
Additional evidence for dynamo action appears in the third panel, where $B_\mathrm{rms}$ grows faster than $\langle\rho\rangle_{\mathrm{J}}^{2/3}$ in runs with $\Rm$ above the critical threshold.
Moreover, the growth rate of $B_\mathrm{rms}$ increases with $\Rey$.
Finally, the bottom panel shows that the ratio $(-W_\mathrm{L} - Q_\mathrm{M})/Q_\mathrm{M}$ is larger for higher $\Rey$.
Here,
$W_\mathrm{L} = \langle \UU \cdot ( \JJ \times  \BB)\rangle$
is the work done by the Lorentz force and
$Q_\mathrm{M} = \langle  \mu_0 \eta \mathbf{J}^2\rangle$ the
Joule dissipation term.
A value of $(-W_\mathrm{L} - Q_\mathrm{M})/Q_\mathrm{M} > 0$ indicates dynamo activity, which we find for all runs in Fig.~\ref{fig_growth_eta} except for Run H1a.
Together, these findings consistently suggest the presence of a small-scale dynamo in high-$\Rey$ runs.

The evolution of the magnetic energy spectrum in runs with different values of $\eta=\nu$ (i.e., ~runs with different Reynolds numbers) is shown in Fig.~\ref{fig_spec_eta}.
In the left panels of Fig.~\ref{fig_spec_eta}, the analysis of Run H1a is shown. 
A forward cascade is evident in $E_\mathrm{M}$, as magnetic energy increasingly populates higher wavenumbers. 
While the spectral peak remains at approximately the same position, the effective correlation wavenumber therefore shifts to larger values, following (with some delay) the value of the position of the Jeans wavenumber $k_\mathrm{J}$ which is indicated by the solid dots in the figure. 
A small-scale dynamo is not expected for Run H1a, as the Reynolds number remains below the critical value for dynamo onset (see the top panel of Fig.~\ref{fig_growth_eta}).
The evolution seen for Run H1a should be similar to that in cosmological simulations in which the Reynolds number is small, and the magnetic field is only amplified by gravitational compression.

In the middle panel of Fig.~\ref{fig_spec_eta}, we show results from Run N1b, which uses $\eta = 10^{-3}$ and has a Reynolds number approximately 40 times higher than that of Run H1a. 
Run N1b shares all parameters with the reference Run H2b, except for the initial magnetic helicity: N1b begins with a nonhelical magnetic field, while H2b is initiated with a helical one. 
Additionally, H2b is run at twice the resolution. A comparison between the middle panels of Fig.~\ref{fig_spec_eta} and Fig.~\ref{fig_spec_gamma} reveals that the forward cascade during gravitational collapse and the onset of the small-scale dynamo seem to be insensitive to the initial magnetic helicity. 
In both N1b and H2b, the spectral growth rate $\gamma_k$ exhibits a $k^{2/3}$ scaling at large $k$ by the final simulation time, indicating the onset of the small-scale dynamo. 

The onset of the small-scale dynamo is most clearly visible in Run H2d, shown in the right panels of Fig.~\ref{fig_spec_eta}. 
In this run, growth at the highest wavenumbers appears in the magnetic energy spectrum $E_\mathrm{M}(k)$ at times $t \gtrsim 2\,t_\mathrm{ff,0}$. 
Although Run H2d does not resolve the resistive scale, which limits the maximum growth rate $\gamma_k$, this rate remains comparable to those measured in Runs N1b and H2b. 
What distinguishes Run H2d, however, is the broader extent of the forward cascade toward higher wavenumbers.
In this case, we observe the emergence of a $E_\mathrm{M}(k) \propto k^{1/3}$ scaling.
This behavior may arise from the forward cascade of magnetic energy, or it may represent an intermediate stage between the initial spectrum and the Kazantsev scaling $E_\mathrm{M}(k) \propto k^{2/3}$ expected during the kinematic phase of the small-scale dynamo.

\section{Discussion }
\label{sec_discussion}

\subsection{Evidence for small-scale dynamo activity}
Overall, the trends presented in this study are consistent with the presence of a small-scale dynamo in simulations with large $\Rey$ values. 
We find that the (time-evolving) values of $B_\mathrm{rms}/\langle \rho \rangle_\mathrm{J}^{2/3}$, the magnetic correlation wavenumber, $k_\mathrm{M}$, and the ratio $(-W_\mathrm{L} - Q_\mathrm{M})/Q_\mathrm{M}$ all increase with the Reynolds number (see Fig.~\ref{fig_growth_eta}). 
Additionally, the wavenumber-dependent growth rate $\gamma_k$ follows a $k^{2/3}$ scaling at late times in simulations where the magnetic Reynolds number exceeds the critical threshold for dynamo action (see lower panels in Figs.~\ref{fig_spec_gamma} and \ref{fig_spec_eta}).
This scaling is consistent with theoretical expectations for magnetic field amplification via the stretching, twisting, and folding of field lines by turbulent eddies, as expressed in Eq.~(\ref{eq_k23}).
Finally, a weak correlation between $B_\mathrm{rms}$ and $\omega_\mathrm{rms}$ is also observed, though this is only visible in the lower-resolution runs which run longer (see Appendix~\ref{sec_correlations}). 

Furthermore, our results suggest that neither the forward cascade nor the early development of the small-scale dynamo depends strongly on the initial magnetic helicity. 
However, it is important to note that our simulations follow the collapse for only a limited time and only capture the onset of the dynamo.
As a result, helicity-dependent effects or other nonlinear behaviors may emerge at later stages, beyond the scope of the current simulations.

\begin{figure}
\centering
    \includegraphics[width=0.4\textwidth]{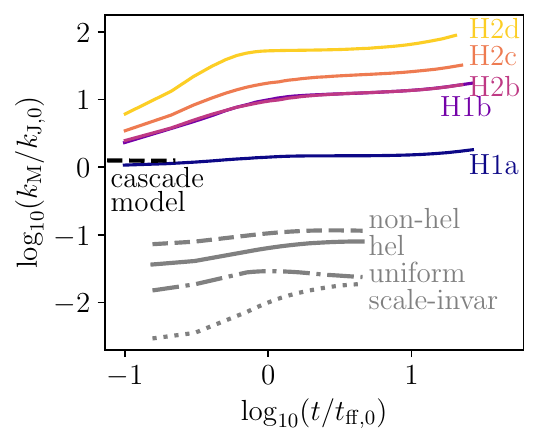}
\caption{Evolution of the magnetic correlation length $k_\mathrm{M}$ for runs with different values of $\eta$ (Runs H1a, H1b, H2b, H2c, H2d, in colored lines. Note: Runs H1b and H2b only differ in resolution) and comparison with two other models. 
In gray colored lines, the results of cosmological simulations by \citet{MtchedlidzeEtAl2022} are shown. 
Different line types indicate different initial magnetic energy spectra (dashed: nonhelical magnetic field, solid: helical magnetic field, dashed-dotted: uniform magnetic field, dotted: scale invariant). 
The black dashed line shows the result of an analytical model of the magnetic forward cascade in gravitational collapse as proposed in \citet{AbramsonEtAl2025}.
The magnetic correlation wavenumber $k_\mathrm{M}$ is normalized by the initial value of the Jeans wavenumber $k_\mathrm{J,0}$ and time is normalized by the initial value of the free-fall time $t_\mathrm{ff,0}$.
Note that Runs H1b and H2b only differ in resolution.}
\label{fig_kM_t}
\end{figure}

\subsection{Comparison with other studies}

We go on to compare our direct numerical simulations of magnetic-field evolution during gravitational collapse with two complementary approaches: (i) the analytical forward-cascade model proposed by \citet{AbramsonEtAl2025}; and (ii) cosmological simulations of large-scale structure formation with different initial magnetic energy spectra, $E_\mathrm{M}$, from \citet{MtchedlidzeEtAl2022}. 
In all three cases, $E_\mathrm{M}$ grows at large $k$ (and in some cases simultaneously at low $k$ as well). 
To enable a quantitative comparison of the spectral evolution, we compute the magnetic-field correlation length in each model and show its time evolution in Fig.~\ref{fig_kM_t}. 
We urge caution, however, as a one-to-one comparison of our results with these studies is not possible; for instance \citet{AbramsonEtAl2025}
have used an idealized model of collapse. In addition, \citet{MtchedlidzeEtAl2022} used a large number of halos, although unresolved (as mentioned above) and their statistics are dominated by the dynamics of magnetic fields on larger scales (i.e., filaments and cosmic voids).

The analytical forward-cascade model by \citet{AbramsonEtAl2025} assumes a constant collapse rate (i.e.~no evolving density) and therefore applies only over a limited time interval. 
Although \citet{AbramsonEtAl2025} show that the spectral peak shifts toward higher $k$, this shift is accompanied by a simultaneous decrease in $E_\mathrm{M}$ at low $k$, such that the resulting value of $k_\mathrm{M}$ changes little (see Fig.~\ref{fig_kM_t}).
Overall, the model is in rough agreement with our Run H1a, in which no turbulence develops.

For the cosmological simulations by \citet{MtchedlidzeEtAl2022}, we show results starting at $z=10$ since this is when 
structures start to form
in those simulations. 
We normalize $k_\mathrm{M}$ by the Jeans wavenumber at $z=10$ and we normalize time by the free-fall time at $z=10$. 
These simulations 
are aimed at understanding magnetic field dynamics on larger scales
whereas our collapse simulations are explicitly designed to include $k_\mathrm{J}$ within the computational domain, 
to understand how turbulence affects field dynamics within collapsing regions.
As a consequence, our runs reach significantly larger values of $k_\mathrm{M}/k_{\mathrm{J},0}$ in Fig.~\ref{fig_kM_t}. \citet{MtchedlidzeEtAl2022} present four different initial magnetic energy spectra motivated by early-Universe magnetogenesis scenarios: nonhelical, helical, uniform, and scale-invariant spectra. 
The helical and nonhelical cases most closely resemble the initial condition assumed in the forward-cascade model of \citet{AbramsonEtAl2025}, in the sense that $E_\mathrm{M}$ peaks at intermediate wavenumbers within the numerical domain. 
Consistent with this, $k_\mathrm{M}/k_{\mathrm{J},0}$ remains approximately constant in these cases
at lower redshifts ($z \lesssim 2$ 
which corresponds to $t/t_\mathrm{ff,0}\gtrsim 1$ in Fig.~\ref{fig_kM_t}).
A similar behavior is seen in our Run H1a, where the viscosity and magnetic diffusivity are large, turbulence does not develop, and only a small shift of magnetic energy toward higher $k$ occurs. 
The cosmological simulation 
which is the most comparable to our turbulent collapse simulations is the scale-invariant case, where the initial $E_\mathrm{M}$ peaks at the smallest wavenumber in the domain. 
In our turbulent runs (H1b, H2b, H2c, H2d), we observe an initial increase in $k_\mathrm{M}$
(seen also in the scale-invariant case),
followed by a phase where it remains roughly constant (near the resistive scale), and finally a renewed increase, 
likely reflecting the shift of the inertial range toward higher $k$ during collapse. 
The latter trend is not seen in the scale-invariant cosmological simulations, most likely because the small-scale dynamo is not resolved at their effective Jeans scale.

\subsection{Limitations and scope}

The simulations presented in this work are designed as controlled direct numerical experiments to isolate the physical mechanisms that govern the spectral evolution of PMFs during gravitational collapse. 
As such, several limitations should be kept in mind when interpreting the results.

First, our DNSs follow the collapse only until the onset of strongly supersonic infall, at which point unresolved shocks develop and the simulations terminate. 
Consequently, we focus on the pre-shock phase of collapse and primarily capture the onset of small-scale dynamo amplification rather than its fully developed nonlinear evolution or saturation. 
While the dynamo signatures identified here (growth of $B_\mathrm{rms}/\langle\rho\rangle^{2/3}$, work balance, and scale-dependent growth rates) provide robust evidence for dynamo activity at sufficiently high Reynolds numbers, longer integrations would be required to quantify the subsequent nonlinear transfer of magnetic energy back toward larger scales.

Second, our model employs an idealized collapse configuration (a supercritical 
Lane-Emden density profile 
in a periodic domain) and assumes an approximately isothermal equation of state. 
This setup isolates the interplay between gravitational compression, turbulence driven on the Jeans scale, and magnetic spectral transfer, but it does not capture additional physical ingredients present in realistic cosmological environments, such as continuous accretion, filamentary inflows, dark-matter potentials, cosmological expansion, or radiative feedback. 
We therefore interpret our results primarily in terms of the underlying dimensionless control parameters (notably the Reynolds number and the ratio $t_\mathrm{SSD}/t_\mathrm{ff}$), which should remain relevant across a broad class of collapsing structures.

Finally, our simulations operate in the regime $\Pm \approx 1$, with explicit viscosity and resistivity chosen such that $\nu=\eta$. 
This choice is motivated by numerical constraints and allows a systematic exploration of the Reynolds number, but it does not probe the extreme magnetic Prandtl numbers expected in many astrophysical plasmas.
Nevertheless, the qualitative distinction between forward-cascade-dominated evolution at low $\Rey$ and dynamo-dominated evolution at high $\Rey$ is expected to be robust.

\section{Conclusion and outlook}
\label{sec_conclusion}

In this paper, we present a series of high-resolution direct numerical simulations of magnetized halos collapsing under self-gravity and an analysis of the evolution of the magnetic energy spectrum. Our main findings can be summarized as follows:

\begin{enumerate}
\item Gravitational collapse drives a forward cascade of magnetic energy
when the initial peak of the magnetic energy spectrum, $k_{\mathrm{peak},0}$,
lies at wavenumbers smaller than the initial Jeans wavenumber (i.e., when magnetic energy is concentrated on scales larger than those of the density field),
$k_{\mathrm{J},0}$ (the opposite ordering leads to inverse transfer and decay of the magnetic field).
In the default configuration considered in this work, $k_{\mathrm{peak},0} < k_{\mathrm{J},0}$, as well as for runs where turbulence does not fully develop due to large dissipation parameters, the magnetic spectrum evolves predominantly through compression and forward spectral transfer. 
In this regime, our results are in approximate agreement with the analytical forward-cascade model of \citet{AbramsonEtAl2025} when considering a simple toy collapse rate and with the cosmological simulations of \citet{MtchedlidzeEtAl2022}, which consider PMFs with different coherence scales.

\item In addition to the forward cascade,
a small-scale dynamo emerges at sufficiently high Reynolds numbers.
Our study shows that whether dynamo amplification becomes dynamically important is determined by the competition between the dynamo growth timescale $t_{\text{SSD}}$ (the inverse growth rate of the small-scale dynamo) and the free-fall time, $t_{\text{ff}}$. 
If $t_{\text{SSD}} \ll t_{\text{ff}}$, the dynamo efficiently amplifies magnetic energy on the viscous scale during collapse. 
If $t_{\text{SSD}} \gg t_{\text{ff}}$, dynamo action is too slow to compete with gravitational contraction and magnetic amplification is dominated by compression and forward spectral transfer.

\item Dynamo action modifies 
(and can even erase) 
primordial spectral features on small length scales. In particular, the small-scale dynamo operates predominantly below the Jeans scale, where turbulence is efficiently generated during gravitational collapse.
Therefore, only magnetic structures on sufficiently large scales can retain memory of PMFs, while smaller scales are dynamically regenerated during structure formation.
\end{enumerate}

Our results suggest that cosmological MHD simulations, which do not resolve the Jeans scale (and therefore do not capture the turbulent inertial range associated with gravitational collapse) are likely to underestimate magnetic-field amplification due to the absence of the small-scale dynamo on small scales. 
In such simulations, the evolution of PMFs may appear dominated by compression and large-scale advection, while the rapid generation of magnetic energy on sub-Jeans scales remains unresolved. 
However, the gravitationally driven forward cascade of magnetic energy toward smaller scales is still captured, even if the small-scale dynamo itself is not fully resolved. 
Consequently, large-scale spectral shifts may be reproduced, whereas the additional dynamo-driven amplification and restructuring of the magnetic spectrum occurs on scales up to the Jeans scale.

More broadly, our findings have important implications for interpreting future measurements of magnetic fields in 
the large-scale structure. In collapsing environments, turbulent dynamo action can quickly reshape the magnetic energy spectrum and erase primordial spectral features on small scales. 
Therefore, accurately linking observed magnetic statistics to primordial initial conditions requires simulations that resolve the turbulent cascade and the relevant dissipation scales, enabling a solid treatment of the competition between gravitationally driven forward transfer and dynamo amplification.

The simulations presented here represent a first step toward disentangling the coupled effects of gravitational collapse and turbulent amplification on the evolution of PMFs in the late Universe. 
Follow-up studies (e.g., idealized galaxy simulations embedded in a cosmological framework) are needed to more precisely quantify the scales on which dynamo amplification dominates magnetic-field growth. 
In addition, a broader exploration of initial magnetic-field configurations (e.g., varying the coherence scale relative to the collapsing region, building on such studies as the works of \citet{FederrathEtAl2011} and \citet{SetaFed2020})
will be essential to assess how the initial conditions influence the onset, efficiency, and characteristic scales of dynamo action.

\begin{acknowledgements}
We acknowledge helpful discussions with Axel Brandenburg, Ralf Klessen,
Hoang (Nhan) Luu, Philip Mocz, Amit Seta, and Mark Vogelsberger. 
The authors gratefully acknowledge access to the Marvin cluster of the University of Bonn. 
SMt acknowledges usage of computational resources on Norddeutscher Verbund f\"ur Hoch- und H\"ochstleistungsrechnen (Germany) and Cineca (Italy; ``IsB30\_MAJIC'').
MA, TK, and SMa acknowledge the National
Science Foundation (NSF) Astronomy and Astrophysics
Research Grants (AAG) Awards AST2307698 and AST2408411; 
TK acknowledges the NASA Astrophysics Theory Program (ATP) Award 80NSSC22K0825; 
TK and SMt also acknowledge the Shota Rustaveli National Science Foundation (SRNSF) of Georgia, FR24-
2606. 
JS, TK, and SMt thank the Bernoulli Center in Lausanne for their hospitality and the interesting discussions during the program ``Generation, evolution, and observations of cosmological magnetic fields''
held in April-June, 2024.
\end{acknowledgements}

\section*{Code and data availability}
In this study, we used the publicly available 
\textsc{Pencil Code} \citep{PencilCodeCollaboration2021}, and Matplotlib \citep{Hunter:2007}.
Selected data files of the simulations presented in this paper are freely available at \href{https://github.com/JenSchober/publications/tree/master/2026/SchoberEtAl_AA}{this repository}.

\begin{appendix}
\nolinenumbers

\onecolumn

\section{Dependence on resolution}
\label{sec_resolution}

Gravitational collapse in the simulations presented in this
work is caused by self-gravity of an overdense region in the 
center of the numerical domain. 
Collapse due to self-gravity is driven by the gradient in the gravitational potential, 
which, in turn, is related to the gradient in density. 
In our simulations, the initial density profile is
a solution of the Lane-Emden equation. 
The resulting initial condition is a density profile with 
$\rho_\mathrm{c}$ at the center and a smooth decrease with radius. 
This density gradient becomes highest at approximately one-half of
the radius.

Higher resolution allows for a more accurate representation of the density gradient, particularly at half the box radius, leading to a more rapid increase in radial velocity. 
Consequently, higher-resolution simulations exhibit an earlier onset of supersonic shocks, ultimately causing the collapse to proceed more quickly and leading to an earlier disruption of the system.

\begin{figure}[h!]
\centering
    \includegraphics[width=0.82\textwidth]{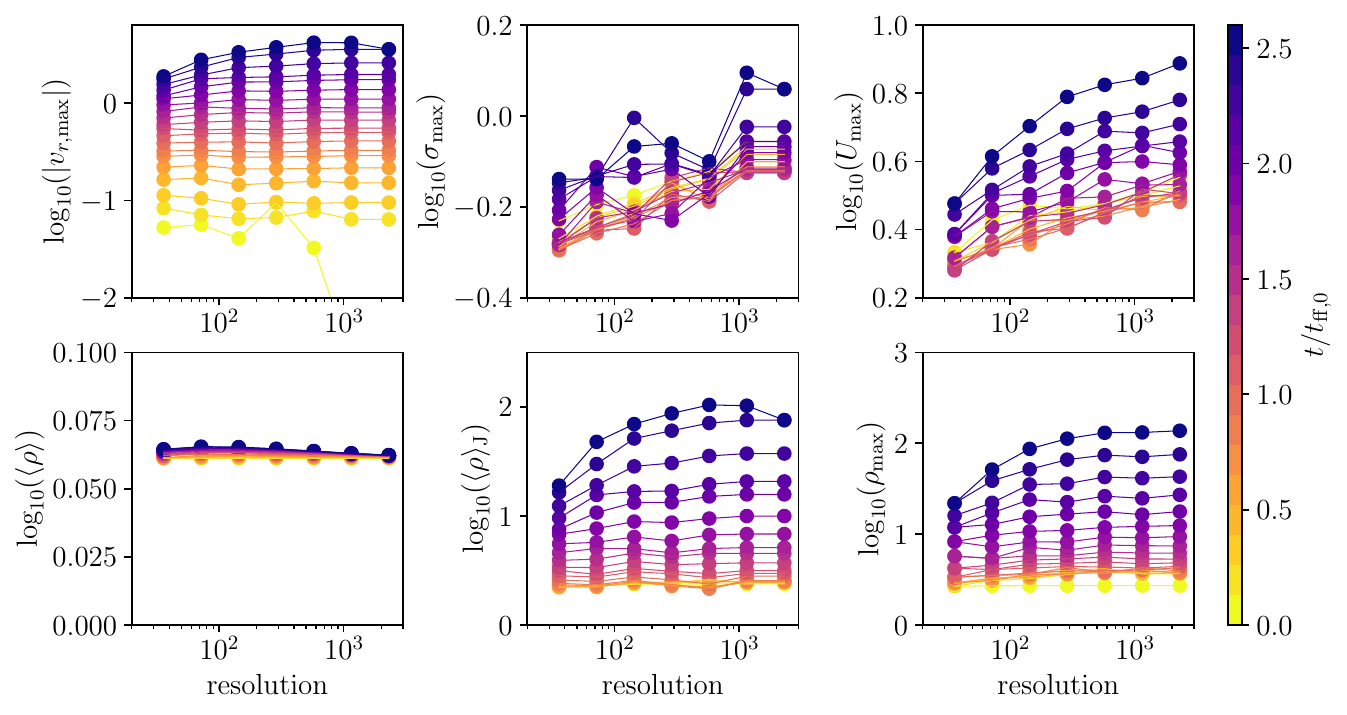}
\caption{Different quantities as a function of time for different resolution. The two runs with the highest resolution shown in this figure are also presented in Tab.~\ref{tab_DNSoverview}
as Runs H1b and H2b.}
\label{fig_v_res}
\end{figure}

In Fig.~\ref{fig_v_res}, we show different quantities related to the gravitational collapse as a function of resolution and for different times. 
In particular, both the spatial maximum of the velocity dispersion $\sigma_\mathrm{max}$ and the spatial maximum of the total velocity  $U_\mathrm{max}$ increase strongly with increasing resolution. 
This leads to a faster development of shocks in high-resolution runs compared to runs with low resolution.

\section{Comparison of timescales for runs with different Reynolds numbers}
\label{sec_timescales}

\begin{figure}[h!]
\centering
 \includegraphics[width=0.82\textwidth]{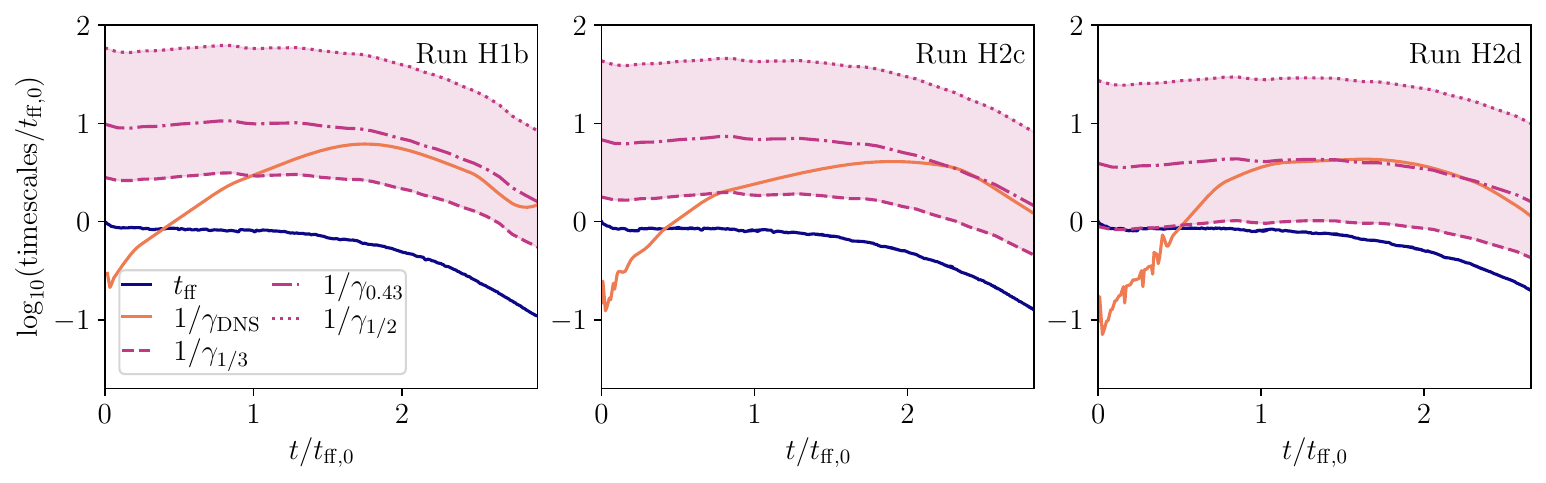}
\caption{Same as upper panel in Fig.~(\ref{fig_growth_t}), but for runs with different Reynolds numbers.}
\label{fig_timescales_t}
\end{figure}

In Fig.~\ref{fig_timescales_t} the time evolution of characteristic timescales is shown for runs with different Reynolds numbers.
Run H2d is closest to the regime where small-scale dynamo dominates ($t_\mathrm{SSD} < t_\mathrm{ff}$), but here the resistive scale is not resolved. 

\section{Run with initial small-scale magnetic fluctuations}
\label{sec_H2bd}

In this appendix, we present Run H2b', in which the initial magnetic energy spectrum is peaked at $k > k_\mathrm{J,0}$. 
A further distinctive feature of this run is that the initial magnetic energy is approximately $1/20$ of the initial kinetic energy, which is significantly higher than in all other runs considered in this study.
For this initial condition, gravitational compression is not expected to amplify the magnetic field, since the initial magnetic energy on scales $k < k_\mathrm{J,0}$ is initially negligible.
The purpose of this run is to investigate whether a small-scale dynamo can develop more efficiently when strong magnetic fluctuations are already present at large wavenumbers.

\begin{figure}[h!]
\centering
    \includegraphics[width=0.35\textwidth]{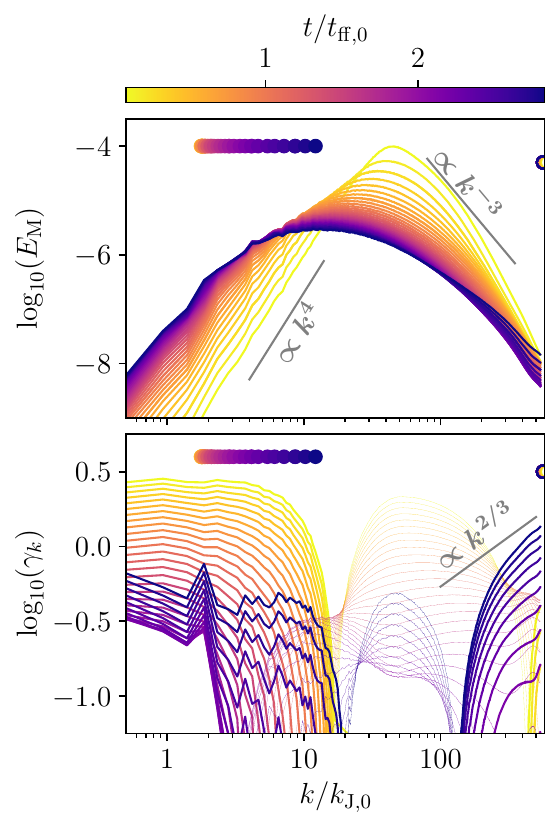}
    \hspace{1cm}
    \includegraphics[width=0.35\textwidth]{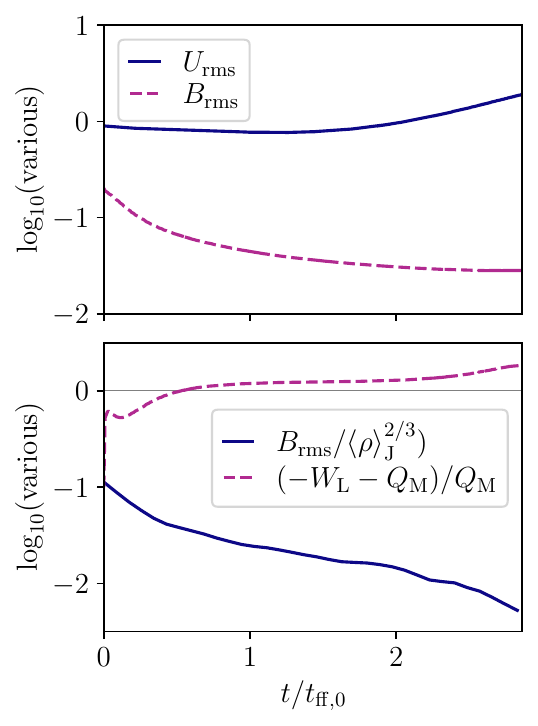}
\caption{Analysis of Run H2b'. Left: Magnetic energy spectrum \textit{(top)} and scale-dependent growth rate of the magnetic energy \textit{(bottom)}. Thick lines indicate that the growth rate is positive, while thin lines (approximately from $k=20$ to $k=200$) indicate that the growth rate is negative and what is shown is its absolute value.
Right: Time evolution of various quantities.
}
\label{fig_H2bd}
\end{figure}

Figure~\ref{fig_H2bd} shows the analysis of run H2b'. 
As anticipated, we do not observe a forward cascade.
Instead, the magnetic energy $E_\mathrm{M}$ decays rapidly at high $k$, 
and the peak of the spectrum $k_\mathrm{peak}$ shifts to smaller wavenumbers. 
This resembles an inverse cascade.
At the final time of Run H2b', the peak of $E_\mathrm{M}$ coincides with $k_\mathrm{J}$.
Interestingly, this decay
of magnetic energy on the initial peak of $E_\mathrm{M}$
is similar to the trend seen in the cosmological simulations of \citet{MtchedlidzeEtAl2022} (see their Figure~6 for helical and nonhelical cases; however, as structures start to form, the power spectrum shows forward cascade, while insignificant growth on larger scales
remains), even though there $k_\mathrm{J,0}\gg k_\mathrm{peak,0}$ (and magnetic field growth in their case is also dominated by the perturbations in the total density field); whereas in our DNSs, we have $k_\mathrm{J,0}< k_\mathrm{peak,0}$.
While in Run H2b' there is some buildup of magnetic energy on scales $k < k_\mathrm{peak}$, the overall $B_\mathrm{rms}$ decreases throughout the entire simulation, as can be seen on the right upper panel of Fig.~\ref{fig_H2bd}.

At $t \approx 2\, t_\mathrm{ff,0}$, the growth rate of the magnetic energy at $k \gtrsim 150$ becomes positive again (see the lower left panel of Fig.~\ref{fig_H2bd}). 
This may indicate the onset of a small-scale dynamo. However, we cannot confirm this interpretation, as there is no corresponding transition to increasing $B_\mathrm{rms}$ or to an increasing ratio $B_\mathrm{rms}/\langle \rho \rangle_\mathrm{J}^{2/3}$ (see the lower right panel of Fig.~\ref{fig_H2bd}).
The only other indication of the presence of a  small-scale dynamo is seen in the evolution of 
$(-W_\mathrm{L}-Q_\mathrm{M})/Q_\mathrm{M}$,
which becomes slightly larger than zero for $t\gtrsim0.6\, t_\mathrm{ff,0}$ (see discussion in Appendix~\ref{sec_work_terms}).

\section{Signatures of the dynamo in the evolution of work terms}
\label{sec_work_terms}

In this appendix, we analyze the potential dynamo activity
by comparing the different work components in the flow,
following the analysis presented in \citet{BrandenburgNtormousi2022}.

\begin{figure}[h!]
\centering
    \includegraphics[width=0.82\textwidth]{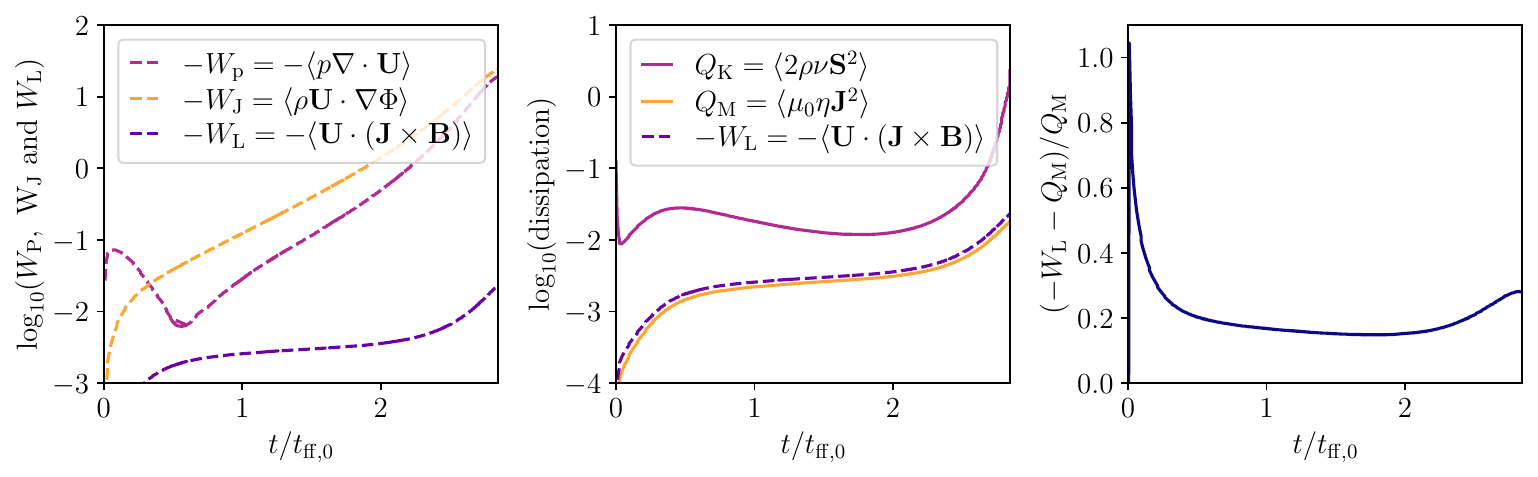}
\caption{Work analysis for Run H2b.
Left: Time evolution of the
work done by the Lorentz force, $W_\mathrm{L}$, 
the pressure force, $W_\mathrm{p}$, and
the gravity term, $W_\mathrm{J}$. are shown.
Middle: $W_\mathrm{L}$ is compared to the 
Joule dissipation term,
$Q_\mathrm{M}$, and the viscous term,
$Q_\mathrm{K}$. Right: $(-W_\mathrm{L}-Q_\mathrm{M})/Q_\mathrm{M}$.
}
\label{fig_work_t}
\end{figure}

In the left panel of Fig.~\ref{fig_work_t}, we compare the time evolution of the
work done by the Lorentz force,
$W_\mathrm{L} = \langle \UU \cdot ( \JJ \times  \BB)\rangle$,
the work done by the pressure force,
$W_\mathrm{p} = - \langle \UU \cdot \nabla p \rangle =  \langle p \nabla \cdot \UU \rangle$,
and the work done by the gravity term,
$W_\mathrm{J} = - \langle \rho \mathbf{U} \cdot \nabla \Phi \rangle$.
All of these work terms are negative, and the absolute values of $W_\mathrm{J}$ and $W_\mathrm{p}$ are greater than that of $W_\mathrm{L}$.
However, $W_\mathrm{L}$ is slightly larger than the Joule dissipation term 
$Q_\mathrm{M} = \langle  \mu_0 \eta \mathbf{J}^2\rangle$, as can be seen in the middle panel of Fig.~\ref{fig_work_t}. 
Viscous dissipation, $Q_\mathrm{K} = \langle 2 \rho \nu \mathbf{S}^2\rangle$, is initially approximately two orders of magnitude greater than 
$Q_\mathrm{M}$, and after $\approx 1\,t_\mathrm{ff,0}$ the difference is reduced to one order of magnitude. 

The right panel of Fig.~\ref{fig_work_t} shows the time evolution of $(-W_\mathrm{L}-Q_\mathrm{M})/Q_\mathrm{M}$.
A value of $(-W_\mathrm{L}-Q_\mathrm{M})/Q_\mathrm{M}>0$, which we find here throughout the simulation, indicates that more magnetic energy is produced than dissipated, thus pointing to dynamo activity.

\section{Correlation between the magnetic and velocity fields}
\label{sec_correlations}

The small-scale dynamo amplifies the magnetic field through stretching, twisting, folding, and merging of field lines by turbulent eddies. 
It should be most efficient in turbulence that is dominated by vorticity, and the magnetic field grows fastest in regions where vorticity is significant. 
If the dynamo is present, we therefore expect a correlation between $B_\mathrm{rms}$ and $\omega_\mathrm{rms} = (\nabla \times \UU)_\mathrm{rms}$.

\begin{figure}[h!]
\centering
    \includegraphics[width=0.49\textwidth]{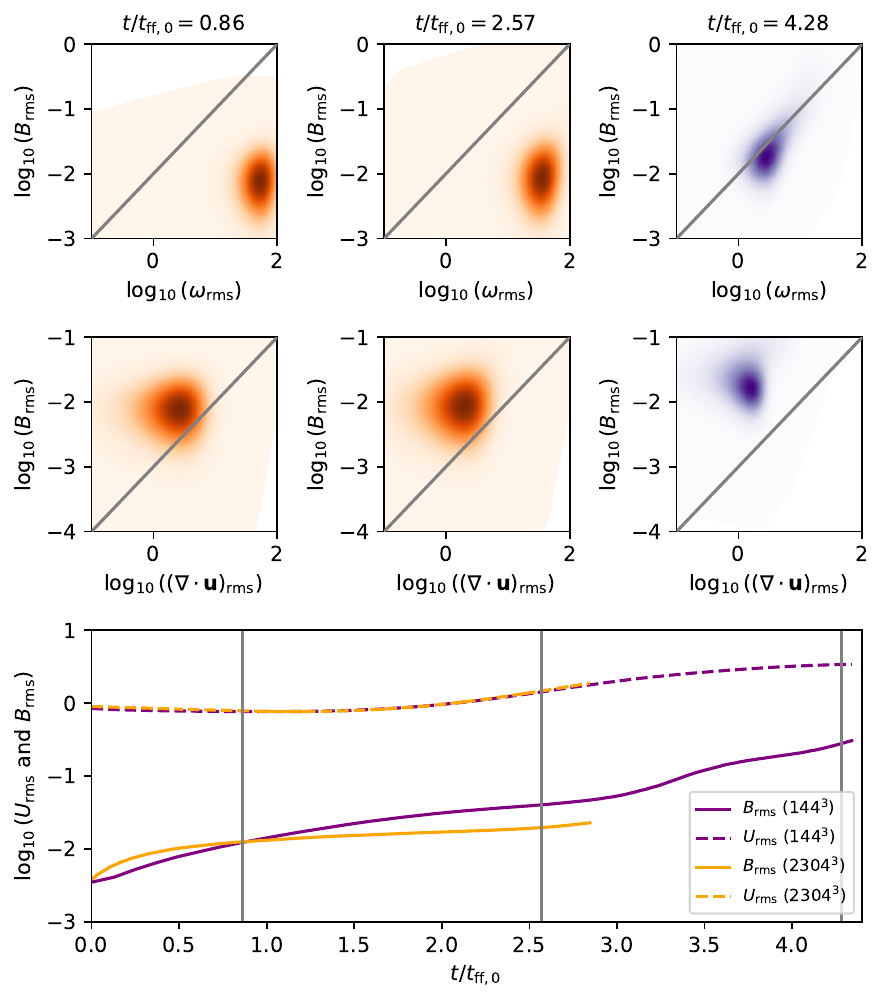}
\caption{Comparison between a high-resolution run (Run H2b, $2304^3$, orange color) with a low-resolution ($144^3$, blue color) with exactly the same parameters.
Top: Cell-wise rms magnetic field strength $B_\mathrm{rms}$ versus cell-wise rms vorticity $\omega_\mathrm{rms} = (\nabla \times \UU)_\mathrm{rms}$ at different times (from left to right) in the 3D simulation domain.
Middle: Cell-wise $B_\mathrm{rms}$ versus the rms divergence of the velocity field, $(\nabla \cdot \UU)_\mathrm{rms}$, again at different times (from left to right).
Bottom: Time evolution of the volume-averaged $U_\mathrm{rms}$ and $B_\mathrm{rms}$ for both runs. 
Vertical gray lines mark the times corresponding to the correlation plots shown above.}
\label{fig_correlations}
\end{figure}

We plot the correlation between $B_\mathrm{rms}$ and $\omega_\mathrm{rms}$ at different times for Run H2b in Fig.~\ref{fig_correlations} in orange color.
As this simulation crashes early, a correlation between $B_\mathrm{rms}$ and $\omega_\mathrm{rms}$ never develops.

As mentioned in Appendix~\ref{sec_resolution}, higher resolution leads to earlier crashes, as the collapse proceeds faster. 
Run H2b has a resolution of $2304^3$, and we compare the correlations in this run with a run that has the same parameters and initial conditions, but a resolution of $144^3$. 
The result for the  $144^3$ run is presented in Fig.~\ref{fig_correlations} in blue color. 
The latter simulation runs longer, and we see that a correlation between 
$B_\mathrm{rms}$ and $\omega_\mathrm{rms}$ has developed at $t= 4.28~t_\mathrm{ff,0}$, as seen in the upper right panel. The extent to which the buildup of a correlation between $B_\mathrm{rms}$ and $\omega_\mathrm{rms}$ is a sign of a small-scale dynamo should be tested in future idealized simulations.

\end{appendix}

\end{document}